\documentclass[twocolumn]{aastex631}

\newcommand{\sub}[1]{_\mathrm{#1}}

\newcommand{\Rp}{R\sub{p}}
\newcommand{\Kpp}{k\sub{2,p}}
\newcommand{\Qp}{Q\sub{p}}
\newcommand{\Kss}{k\sub{2,\star}}
\newcommand{\Qs}{Q_\star}
\newcommand{\koQp}{\Kpp/\Qp}
\newcommand{\koQs}{\Kss/\Qs}
\newcommand{\porb}{P\sub{orb}}
\newcommand{\prot}{P\sub{rot\_ini, \star}}
\newcommand{\eps}{\varepsilon\sub{\star}}
\newcommand{\gyrs}{\zeta\sub{\star}}

\newcommand{\unit}[1]{\ensuremath{\, \mathrm{#1}}} %Remove if necessary

\providecommand{\msun}{\ensuremath{\,M_\Sun}}
\providecommand{\rsun}{\ensuremath{\,R_\Sun}}

\newcommand{\PSUAA}{Department of Astronomy \& Astrophysics, 525 Davey Laboratory, The Pennsylvania State University, University Park, PA 16802, USA}
\newcommand{\PSUCEHW}{Center for Exoplanets and Habitable Worlds, 525 Davey Laboratory, The Pennsylvania State University, University Park, PA 16802, USA}
\newcommand{\PSETI}{Penn State Extraterrestrial Intelligence Center, 525 Davey Laboratory, The Pennsylvania State University, University Park, PA 16802, USA}
\newcommand{\NOAO}{NSF's National Optical-Infrared Astronomy Research Laboratory, 950 N.\ Cherry Ave., Tucson, AZ 85719, USA}

\usepackage[T1]{fontenc} % for Icelandic Characters
\usepackage[utf8]{inputenc} % for Icelandic Characters 

\usepackage{ulem} % strikethrough with \sout{}

\received{25 January 2023}
\revised{2 June 2023}
\accepted{16 June 2023}

\submitjournal{AJ}

\shorttitle{TOI-1899 update}
\shortauthors{Lin et al.}

\graphicspath{{./}{figures/}}

\begin{document}

\title{The unusual M-dwarf Warm Jupiter TOI-1899~b: Refinement of orbital and planetary parameters}

\correspondingauthor{Andrea S.J.\ Lin}
\email{asjlin@psu.edu}

\author[0000-0002-9082-6337]{Andrea S.J.\ Lin}
\affil{\PSUAA}
\affil{\PSUCEHW}

\author[0000-0002-2990-7613]{Jessica E. Libby-Roberts}
\affil{\PSUAA}
\affil{\PSUCEHW}

\author[0000-0003-0353-9741]{Jaime A. Alvarado-Montes}
\affil{School of Mathematical and Physical Sciences, Macquarie University, Balaclava Road, North Ryde, NSW 2109, Australia}
\affil{The Macquarie University Astrophysics and Space Technologies Research Centre, Macquarie University, Balaclava Road, North Ryde, NSW 2109, Australia}

\author[0000-0003-4835-0619]{Caleb I. Ca\~nas}
\altaffiliation{NASA Postdoctoral Fellow}
\affil{NASA Goddard Space Flight Center, 8800 Greenbelt Road, Greenbelt, MD 20771, USA}
\affil{\PSUAA}
\affil{\PSUCEHW}

\author[0000-0001-8401-4300]{Shubham Kanodia}
\affil{Earth and Planets Laboratory, Carnegie Institution for Science, 5241 Broad Branch Road, NW, Washington, DC 20015, USA} 

\author[0000-0002-7127-7643]{Te Han}
\affil{Department of Physics \& Astronomy, The University of California, Irvine, Irvine, CA 92697, USA}

% observers
\author[0000-0003-1263-8637]{Leslie Hebb}
\affil{Physics Department, Hobart and William Smith Colleges, 300 Pulteney Street, Geneva, NY 14456, USA}
\affil{Department of Astronomy, Cornell University, 245 East Ave, Ithaca, NY 14850, USA}

\author[0000-0002-4625-7333]{Eric L.\ N.\ Jensen}
\affil{Department of Physics \& Astronomy, Swarthmore College, Swarthmore, PA 19081, USA}

\author[0000-0001-9596-7983]{Suvrath Mahadevan}
\affil{\PSUAA}
\affil{\PSUCEHW}
\affil{ETH Zurich, Institute for Particle Physics \& Astrophysics, Zurich, Switzerland}

\author[0000-0002-5300-5353]{Luke C. Powers}
\affil{\PSUAA}

\author[0000-0002-5817-202X]{Tera N. Swaby}
\affil{Department of Physics \& Astronomy, University of Wyoming, 1000 E.\ University, Dept 3905, Laramie, WY 82071, USA}

\author[0000-0001-9209-1808]{John Wisniewski}
\affil{NASA Headquarters, 300 Hidden Figures Way SW, Washington, DC 20546, USA}

% alphabetical

\author[0000-0001-7708-2364]{Corey Beard}
\affil{Department of Physics \& Astronomy, The University of California, Irvine, Irvine, CA 92697, USA}

\author[0000-0003-4384-7220 ]{Chad F. Bender}
\affil{Steward Observatory, University of Arizona, 933 N Cherry Ave, Tucson, AZ 85721, USA}

\author[0000-0002-6096-1749]{Cullen H. Blake}
\affil{Department of Physics and Astronomy, University of Pennsylvania, 209 South 33rd Street, Philadelphia, PA 19104, USA}

\author[0000-0001-9662-3496]{William D. Cochran}
\affil{McDonald Observatory and Center for Planetary Systems Habitability, The University of Texas, Austin, TX, 78712, USA}

\author[0000-0002-2144-0764]{Scott A. Diddams}
\affil{Electrical, Computer \& Energy Engineering, 425 UCB, University of Colorado, Boulder, CO 80309, USA}
\affil{Department of Physics, 390 UCB, University of Colorado, Boulder, CO 80309, USA}
\affil{National Institute of Standards \& Technology, 325 Broadway, Boulder, CO 80305, USA}

\author[0000-0001-6569-3731]{Robert C. Frazier}
\affil{\PSUAA}
\affil{\PSUCEHW}

\author[0000-0002-0560-1433]{Connor Fredrick}
\affil{National Institute of Standards \& Technology, 325 Broadway, Boulder, CO 80305, USA}
\affil{Department of Physics, 390 UCB, University of Colorado, Boulder, CO 80309, USA}

\author[0000-0002-4020-3457]{Michael Gully-Santiago}
\affil{Department of Astronomy, The University of Texas at Austin, Austin, TX, 78712, USA}

\author[0000-0003-1312-9391]{Samuel Halverson}
\affil{Jet Propulsion Laboratory, 4800 Oak Grove Drive, Pasadena, CA 91109, USA}

\author[0000-0002-9632-9382]{Sarah E. Logsdon}
\affil{\NOAO}

\author[0000-0003-0241-8956]{Michael W.\ McElwain}
\affil{Exoplanets and Stellar Astrophysics Laboratory, NASA Goddard Space Flight Center, Greenbelt, MD 20771, USA}

\author[0000-0002-4404-0456]{Caroline Morley}
\affil{Department of Astronomy, The University of Texas at Austin, Austin, TX, 78712, USA}

\author[0000-0001-8720-5612]{Joe P.\ Ninan}
\affil{Department of Astronomy and Astrophysics, Tata Institute of Fundamental Research, Homi Bhabha Road, Colaba, Mumbai 400005, India}

\author[0000-0002-2488-7123]{Jayadev Rajagopal}
\affil{\NOAO}

\author[0000-0002-4289-7958]{Lawrence W. Ramsey}
\affil{\PSUAA}
\affil{\PSUCEHW}

\author[0000-0003-0149-9678]{Paul Robertson}
\affil{Department of Physics \& Astronomy, The University of California, Irvine, Irvine, CA 92697, USA}

\author[0000-0001-8127-5775]{Arpita Roy}
\affil{Space Telescope Science Institute, 3700 San Martin Drive, Baltimore, MD 21218, USA}
\affil{Department of Physics and Astronomy, Johns Hopkins University, 3400 N Charles St, Baltimore, MD 21218, USA}

\author[0000-0002-4046-987X]{Christian Schwab}
\affil{School of Mathematical and Physical Sciences, Macquarie University, Balaclava Road, North Ryde, NSW 2109, Australia}

\author[0000-0001-7409-5688]{Gu{\dh}mundur Stef{\'a}nsson}
\altaffiliation{NASA Sagan Fellow}
\affil{Department of Astrophysical Sciences, Princeton University, 4 Ivy Lane, Princeton, NJ 08540, USA}

\author[0000-0002-5951-8328]{Daniel J. Stevens}
\affil{Department of Physics \& Astronomy, University of Minnesota Duluth, Duluth, MN 55812, USA}

\author[0000-0002-4788-8858]{Ryan C. Terrien}
\affil{Carleton College, One North College St., Northfield, MN 55057, USA}

\author[0000-0001-6160-5888]{Jason T.\ Wright}
\affil{\PSUAA}
\affil{\PSUCEHW}
\affil{\PSETI}


\begin{abstract}

TOI-1899~b is a rare exoplanet, a temperate Warm Jupiter orbiting an M-dwarf, first discovered by \citet{Canas2020_toi1899} from a TESS single-transit event. Using new radial velocities (RVs) from the precision RV spectrographs HPF and NEID, along with additional TESS photometry and ground-based transit follow-up, we are able to derive a much more precise orbital period of $P = 29.090312_{-0.000035}^{+0.000036}$~d, along with a radius of $R_p = 0.99 \pm 0.03~R_J$. We have also improved the constraints on planet mass, $M_p = 0.67 \pm 0.04~M_J$, and eccentricity, which is consistent with a circular orbit at 2$\sigma$ ($e = 0.044_{-0.027}^{+0.029}$). TOI-1899~b occupies a unique region of parameter space as the coolest known ($T_{eq} \approx$ 380~K) Jovian-sized transiting planet around an M-dwarf; we show that it has great potential to provide clues regarding the formation and migration mechanisms of these rare gas giants through transmission spectroscopy with JWST as well as studies of tidal evolution.

\end{abstract}


\keywords{radial velocity, transit photometry, extrasolar gaseous planets}

\section{Introduction}
\label{sec:intro}

% M-dwarf planets, esp. gas giants, are rare -- makes sense with lower disk mass, but we DO see a few. How did they form? Also not well followed-up / characterized
% TESS followup -- single transits are tricky, especially with long duration transits
% are M dwarf formation channels different? tidal forces act over longer timescales

The vast majority of planets discovered around M-dwarfs are small, with gas giants ($R_p > 8~R_\oplus$) being rare despite being easier to detect through either transit photometry or radial velocities (RVs). It is relatively easy to explain why such gas giants are thought to be intrinsically rare---both models \citep[e.g.,][]{Laughlin2004} and observations \citep[e.g.,][]{observe.disk.dependence} indicate that M-dwarf protoplanetary disks are much smaller and less massive than disks around more massive FGK stars. With less gas and dust available, M-dwarf disks should be much less likely to form gas giants \citep{Burn2021_formation}, especially via the core-accretion pathway \citep{Ida2004_corecollapse} because the timescale for formation of giant cores is too long relative to the disk lifetime. However, it remains unclear why some M-dwarf disks \textit{are} capable of forming these giant planets, though M-dwarf gas giants seem to form preferentially around metal-rich stars \citep{Maldonado2020_Mdwarfmetallicity, Gan2022_toi530}, which may host more massive disks with large enough quantities of solids for massive planetary cores to form.

With the recent abundance of space-based photometric data from the Transiting Exoplanet Survey Satellite \citep[TESS;][]{Ricker2014_TESS}, we are beginning to find more examples of these rare M-dwarf gas giants, including nine short-period Jupiters ($P \leq 10$~d) (see \citealt{Canas2022_3714_3629}, \citealt{Gan2022_toi530}, and references therein). However, Warm Jupiters (WJs)---here defined as gas giant planets with periods between 10 and 100 days---around M-dwarfs are much rarer, with only three other examples known apart from TOI-1899~b: the two statistically-validated planets Kepler-1628~b \citep[$P$ = 76.4~d, $R_p = 6.5~R_\oplus$;][]{Morton2016_kepler1628} and K2-387~b \citep[$P$ = 28.7~d, $R_p = 7.3~R_\oplus$;][]{Christiansen2022_k2-387}, and the very young TOI-1227~b \citep[$P$ = 27.4~d, $R_p = 9.6~R_\oplus$;][]{Mann2022_toi1227}, which is expected to contract to $\lesssim 5~R_\oplus$ as it cools. However, TOI-1899~b is the only M-dwarf WJ with a mass measurement. WJs are challenging to discover in TESS photometry, as their longer periods often mean they are detected only as single-transit events. Even with precision RV follow-up, having only a single transit means it is often difficult to obtain an ephemeris precise enough to schedule future transit observations (e.g., for transmission spectroscopy). Being able to do so often relies on TESS revisiting the field in later sectors, or extensive reconnaissance observations with ground-based photometry.

Warm Jupiters around FGK stars largely exhibit low to moderate eccentricities, with the few with high eccentricities thought to be ``passing through'' on their way to circularize as Hot Jupiters, or possibly excited by companion-induced eccentricity cycling \citep{Dawson2018_HJ}. Many low-eccentricity WJs also host smaller inner companions \citep[e.g.,][]{Weiss2013_koi94,Tran2022_toi1670}, with \citet{Huang2016_WJcompanions} suggesting the companion fraction is $\sim$50\%. In turn, the high companion fraction indicates that a large fraction of FGK WJs must form through relatively peaceful, dynamically-cool mechanisms, e.g., forming \textit{in situ} or undergoing disk migration. However, since very few M-dwarf WJs are known, it is unknown whether they form in the same ways and at the same rates as their FGK cousins, or whether the small size and low mass of M-dwarf disks compels different formation and evolutionary pathways.

% good from here

The M-dwarf Warm Jupiter TOI-1899~b was discovered from a TESS single-transit event (with two sectors of coverage, Sectors 14 and 15) by \citet{Canas2020_toi1899}, who constrained its orbital period with 15 RVs obtained with the near-infrared (NIR) Habitable-zone Planet Finder spectrograph \citep[HPF;][]{Mahadevan2012_hpf, Mahadevan2014_hpf}. In this paper, we update the parameters and orbital elements of this system, using additional RVs from HPF and NEID, three more TESS transits, and ground-based transit photometry.

Section \ref{sec:obs_data} describes our RV and photometric observations and data processing, and Section \ref{sec:stellar} details what these data tell us about stellar parameters, including activity levels and rotation period. Section \ref{sec:joint_fit} presents the updated parameters of the TOI-1899 system. In Section \ref{sec:discussion}, we detail how TOI-1899~b occupies a unique region of parameter space, which makes it a scientifically interesting target both for further atmospheric characterization with JWST and for tidal evolution simulations.

\section{Observations} % DONE EXCEPT UPDATE FIGURE
\label{sec:obs_data}

\subsection{HPF RVs}
\label{sec:hpf}

HPF is a fiber-fed \citep{Kanodia2018_hpf_fibers}, high-resolution (R $\sim$ 50,000), NIR (808 to 1278~nm) spectrograph. It is thermally stabilized at the milli-Kelvin level \citep{Stefansson2016_hpf_environment}, and has demonstrated an on-sky RV precision of 1.53 m/s \citep{Metcalf2019_hpf_precision}. HPF is located at the 10m Hobby-Eberly Telescope \citep[HET;][]{Ramsey1998_HET, Hill2021_HET} at McDonald Observatory, which is a fixed-altitude telescope with a roving pupil design, and is fully queue-scheduled with all observations executed in a queue by the HET resident astronomers \citep{Shetrone2007_het_queue}.

TOI-1899~b was originally published by \citet{Canas2020_toi1899} with 15 RVs from HPF. Since then, we have obtained 35 more HPF visits, with each visit comprising two 945~s exposures. We discard 6 visits which have signal-to-noise ratios less than half the median S/N-per-pixel of 57 (measured at 1000~nm). The remaining 29 visits are shown in \autoref{tab:rvs}. HPF engineering work in May 2022 required thermal and vacuum cycling of the instrument, resulting in an RV offset once HPF had re-stabilized; we fit separately the RVs taken before and after the velocity break. The phase-folded RVs of TOI-1899 are shown in the top panel of \autoref{fig:joint_fit_rv}.

HPF is capable of simultaneous calibration using a NIR Laser Frequency Comb \citep[LFC;][]{Metcalf2019_hpf_precision}, but we chose not to use simultaneous calibration due to the faintness of our target, in order to minimize the impact of scattered light in the science spectrum. Instead, we obtain a wavelength solution by interpolating the wavelength solution from other LFC exposures on the night of the observations, which has been shown to enable wavelength calibration and drift correction at a precision of $\sim$30 cm/s \citep{Stefansson2020_g940}, a value much smaller than our estimated per-observation RV uncertainty (instrumental precision and photon noise, added in quadrature) of $\sim$15 m/s on this target.

We use \texttt{HxRGproc} \citep{Ninan2018_hpf_ccd} for correction of the raw HPF data, and then derive RVs via the methodology outlined in \citet{Metcalf2019_hpf_precision}. We use a modified version of the \texttt{SpEctrum Radial Velocity AnaLyser} \citep[\texttt{SERVAL};][]{Zechmeister2018_serval} as discussed further in \citet{Stefansson2020_g940}, which employs the template-matching technique to derive RVs \citep[e.g.,][]{AngladaEscude2012_templatematching}. We generated the master template using all observed spectra, ignoring telluric regions identified by using a synthetic telluric-line mask generated from \texttt{telfit} \citep{Gullikson2014_telfit}, a Python wrapper to the Line-by-Line Radiative Transfer Model package \citep{Clough2005_lblrtm}. We calculated the barycentric correction for each epoch using \texttt{barycorrpy} \citep{Kanodia2018_barycorrpy}, the Python implementation of the algorithms from \cite{Wright2014_barycorr}.

\subsection{NEID RVs}
\label{sec:neid}

NEID \citep{Schwab2016_neid, Halverson2016_errorbudget} is an ultra-stabilized \citep{Robertson2019_neid_environment}, high-resolution (R $\sim$ 110,000) spectrograph mounted on the WIYN 3.5m Telescope\footnote{The WIYN Observatory is a joint facility of the NSF's National Optical-Infrared Astronomy Research Laboratory, Indiana University, the University of Wisconsin-Madison, Pennsylvania State University, the University of Missouri, the University of California-Irvine, and Purdue University.} at Kitt Peak National Observatory. NEID is fiber-fed via a dedicated port adapter on the WIYN 3.5m \citep{Schwab2018_neid_port, Logsdon2018_neid_port}, and covers a broad red-optical wavelength range from 380 to 930~nm. It is wavelength-calibrated by an astro-comb (a purpose-built LFC) and a Fabry-P\'erot etalon. Like HPF, NEID is also capable of simultaneous calibration (using the etalon instead of the LFC), but again we chose not to use this feature due to the faintness of our target.

\begin{figure*}[!htbp]
    \centering
    \includegraphics[width=0.6\textwidth]{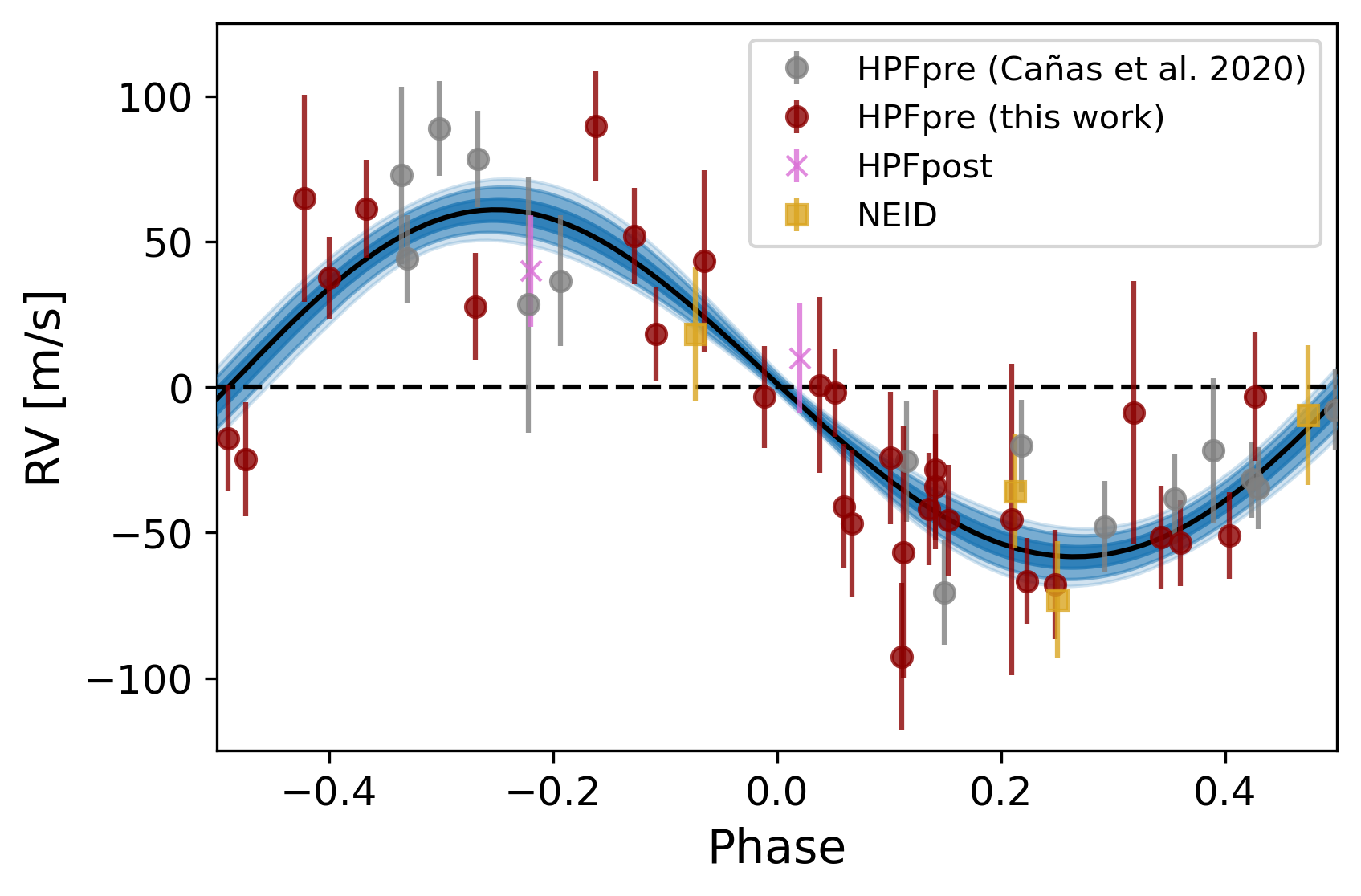}\\
    \bigskip
    \includegraphics[width=\textwidth]{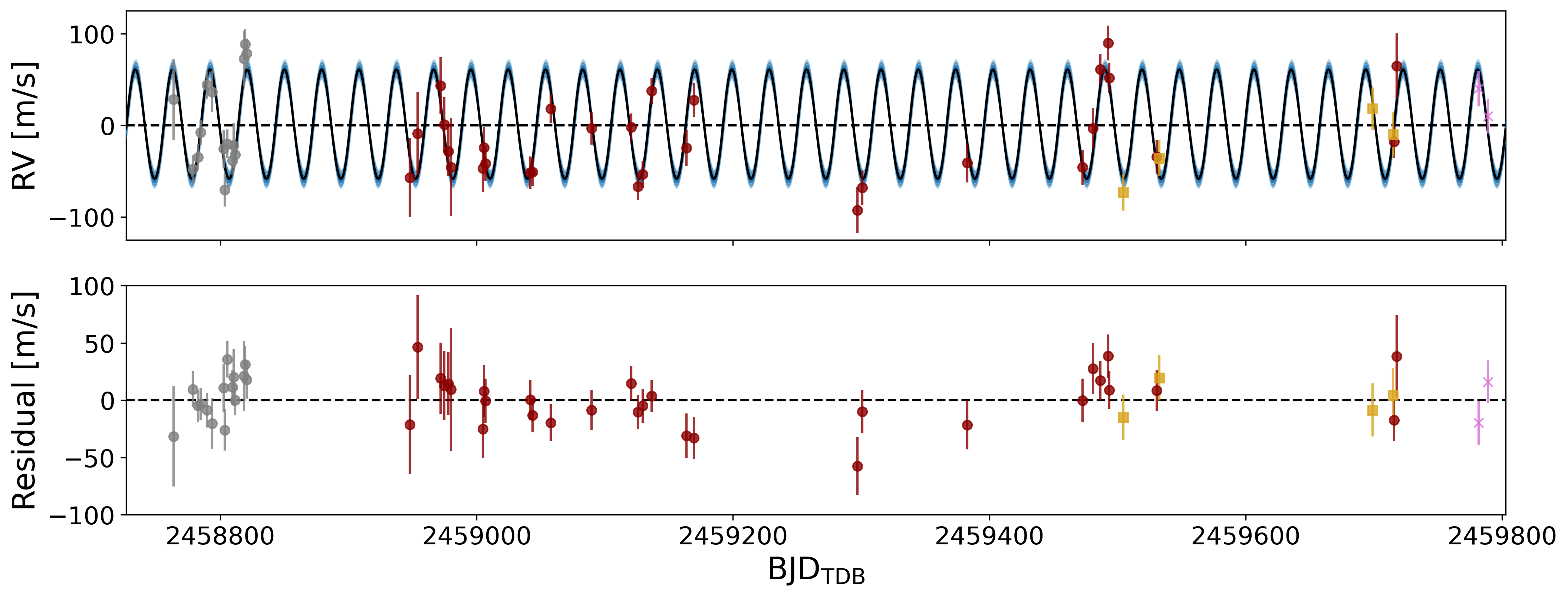}
    \caption{Joint-fit RVs of TOI-1899~b. The best-fit model is plotted as a black line, while the blue shaded regions denote the $1\sigma$ (darkest), $2\sigma$, and $3\sigma$ ranges of the derived posterior solution. \textit{Top}: Phase-folded RVs, with HPF data from \citet{Canas2020_toi1899} in grey, new HPF data in dark red (pre-velocity break) and pink (post-velocity break), and NEID in gold. \textit{Bottom}: Unphased RVs and residuals as a function of time.}
    \label{fig:joint_fit_rv}
\end{figure*}

We also obtained 4 visits of TOI-1899 with NEID in HR (High Resolution) mode, with single exposures of 1800~s per visit. Of our 4 visits, 2 exhibit significantly lower S/N due to poor observing conditions, but we include them anyway due to the small number of NEID RVs. These are shown in \autoref{tab:rvs} and the top panel of \autoref{fig:joint_fit_rv}.

NEID data are automatically reduced by the NEID Data Reduction Pipeline\footnote{\url{https://neid.ipac.caltech.edu/docs/NEID-DRP/}}. We retrieve the Level-2 (fully processed) 1D spectra from the NEID Data Archive\footnote{\url{https://neid.ipac.caltech.edu/search.php}}, and use a NEID-adapted version of \texttt{SERVAL} \citep{Stefansson2022_GJ3470b} to derive the final RVs. These RVs are calculated from 44 orders centered between 5070--8900 $\text{\AA}$ (order indices 52 to 104, corresponding to echelle orders 121 to 69) using only the central-most 3000 pixels of each order, which has been found to result in better RV precision for faint M-dwarfs like this one, by using only the highest S/N portions of the spectra \citep[see e.g.,][]{Canas2022_3714_3629}.

% \startlongtable
\begin{deluxetable}{rrrrl}[!htbp]
\centering
\caption{RVs of TOI-1899}
\label{tab:rvs}
\tabletypesize{\footnotesize}
\tablehead{\colhead{BJD$_{\rm{TDB}}$} & \colhead{RV (m/s)} & \colhead{$\sigma$ (m/s)} & \colhead{S/N\tablenotemark{a}} & Instrument\tablenotemark{b}}
\startdata
2458763.68342 & 52.68  & 43.54 & 34  & HPFpre\tablenotemark{c}  \\
2458778.65399 & -23.56 & 14.51 & 68  & HPFpre  \\
2458782.63085 & -10.48 & 12.82 & 75  & HPFpre  \\
2458784.63162 & 16.50  & 12.63 & 77  & HPFpre  \\
2458789.62112 & 68.47  & 13.83 & 71  & HPFpre  \\
2458793.60354 & 60.85  & 21.72 & 47  & HPFpre  \\
2458802.58906 & -1.13  & 19.94 & 50  & HPFpre  \\
2458803.57261 & -46.29 & 16.95 & 59  & HPFpre  \\
2458805.58078 & 4.14   & 14.73 & 67  & HPFpre  \\
2458809.56162 & -14.05 & 14.30 & 68  & HPFpre  \\
2458810.55741 & 2.57   & 24.09 & 42  & HPFpre  \\
2458811.55496 & -7.49  & 11.77 & 83  & HPFpre  \\
2458818.55514 & 97.10  & 30.01 & 36  & HPFpre  \\
2458819.54226 & 113.14 & 15.21 & 67  & HPFpre  \\
2458820.54718 & 102.72 & 15.44 & 64  & HPFpre  \\
\hline
2458971.88328 & 67.59  & 30.63 & 34  & HPFpre  \\
2458974.88034 & 24.97  & 29.68 & 35  & HPFpre  \\
2458977.88244 & -3.99  & 26.67 & 38  & HPFpre  \\
2459004.80488 & -22.52 & 24.76 & 41  & HPFpre  \\
2459005.79603 & 0.05   & 21.99 & 45  & HPFpre  \\
2459006.79804 & -17.51 & 18.45 & 53  & HPFpre  \\
2459041.93039 & -27.15 & 16.61 & 58  & HPFpre  \\
2459043.70033 & -26.60 & 13.71 & 71  & HPFpre  \\
2459057.88720 & 42.62  & 14.87 & 66  & HPFpre  \\
2459089.80308 & 20.95  & 16.62 & 60  & HPFpre  \\
2459120.72756 & 22.37  & 13.80 & 72  & HPFpre  \\
2459125.70312 & -42.30 & 13.50 & 73  & HPFpre  \\
2459129.69317 & -29.26 & 13.62 & 72  & HPFpre  \\
2459136.67793 & 61.82  & 12.86 & 76  & HPFpre  \\
2459163.59938 & -0.52  & 18.64 & 53  & HPFpre  \\
2459169.57385 & 51.91  & 17.52 & 57  & HPFpre  \\
2459296.99568 & -68.24 & 24.51 & 42  & HPFpre  \\
2459300.99395 & -43.55 & 17.76 & 55  & HPFpre  \\
2459382.77297 & -16.71 & 20.60 & 49  & HPFpre  \\
2459472.74724 & -21.35 & 18.15 & 55  & HPFpre  \\
2459480.72502 & 21.16  & 21.48 & 50  & HPFpre  \\
2459486.71612 & 85.63  & 15.79 & 64  & HPFpre  \\
2459492.69426 & 114.13 & 17.94 & 57  & HPFpre  \\
2459493.69425 & 76.29  & 15.50 & 65  & HPFpre  \\
2459504.68897 &	-74.44 & 7.71  & 16  & NEID \\
2459530.59402 & -9.95  & 17.22 & 58  & HPFpre  \\
2459532.67008 &	-37.32 & 7.08  & 17  & NEID \\
2459698.89831 &	16.71  & 14.02 & 9   & NEID \\
2459714.81917 &	-11.16 & 15.39 & 9   & NEID \\
2459715.86110 & 6.71   & 17.15 & 59  & HPFpre  \\
2459717.84933 & 89.17  & 35.15 & 34  & HPFpre  \\
2459781.91055 & -10.38 & 19.07 & 51  & HPFpost  \\
2459788.89371 & -40.33 & 18.69 & 52  & HPFpost 
\enddata
\tablecomments{Data above the horizontal line were first published in \citet{Canas2020_toi1899}; data below are presented for the first time in this work.}
\tablenotetext{a}{HPF S/N measured in order index 18 ($\sim$1000~nm), NEID S/N in order index 102 ($\sim$850~nm).}
\tablenotetext{b}{``HPFpre'' and ``HPFpost'' indicate data taken before and after the HPF engineering velocity break, respectively.}
\tablenotetext{c}{Exposure time of 945~s.}
\end{deluxetable}

\begin{figure*}[!htbp]
    % transits
    \begin{minipage}[b]{0.5\linewidth}
    \centering
    \includegraphics[width=\textwidth]{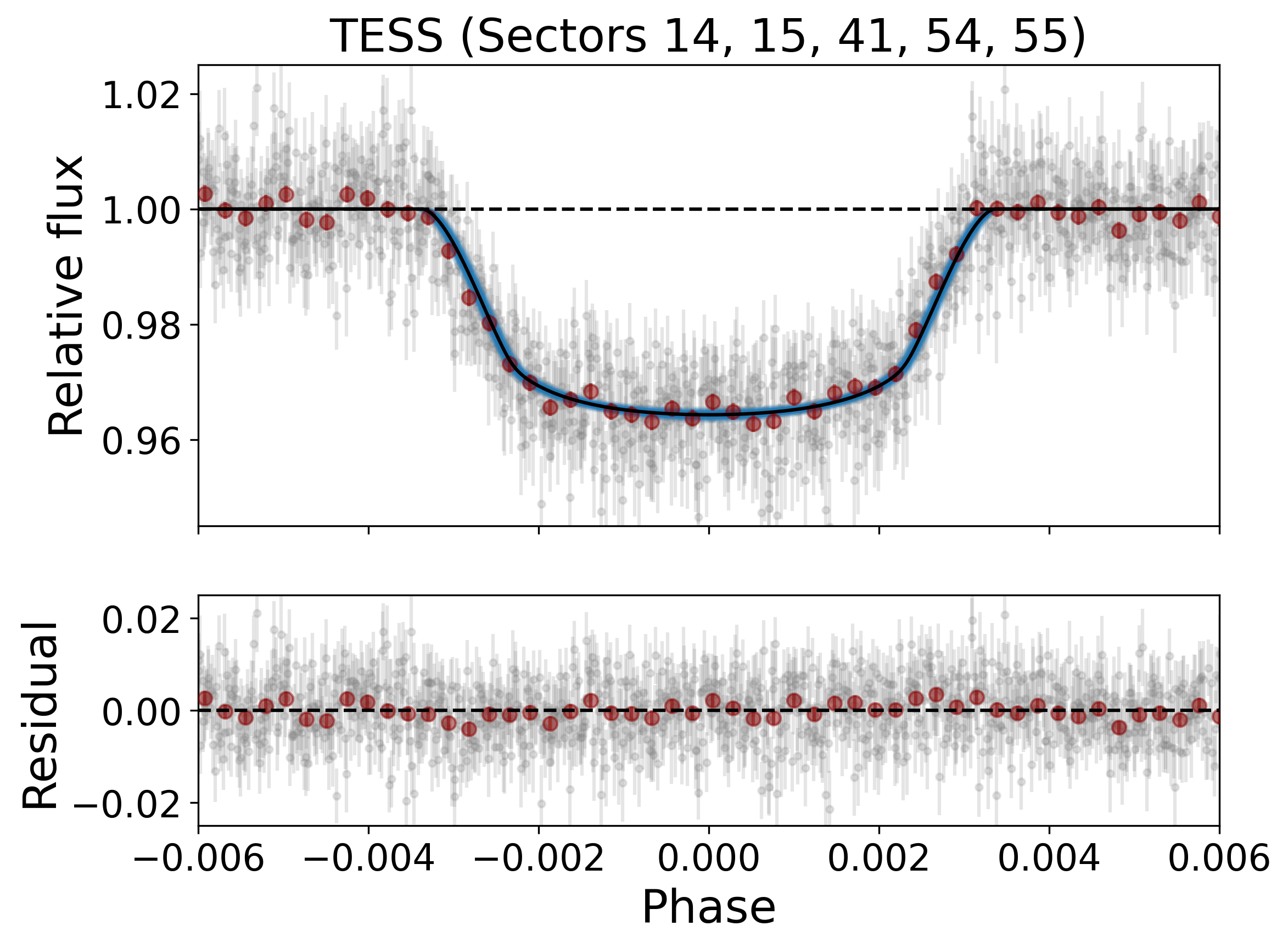}
    \end{minipage}
    \begin{minipage}[b]{0.5\linewidth}
    \centering
    \includegraphics[width=\textwidth]{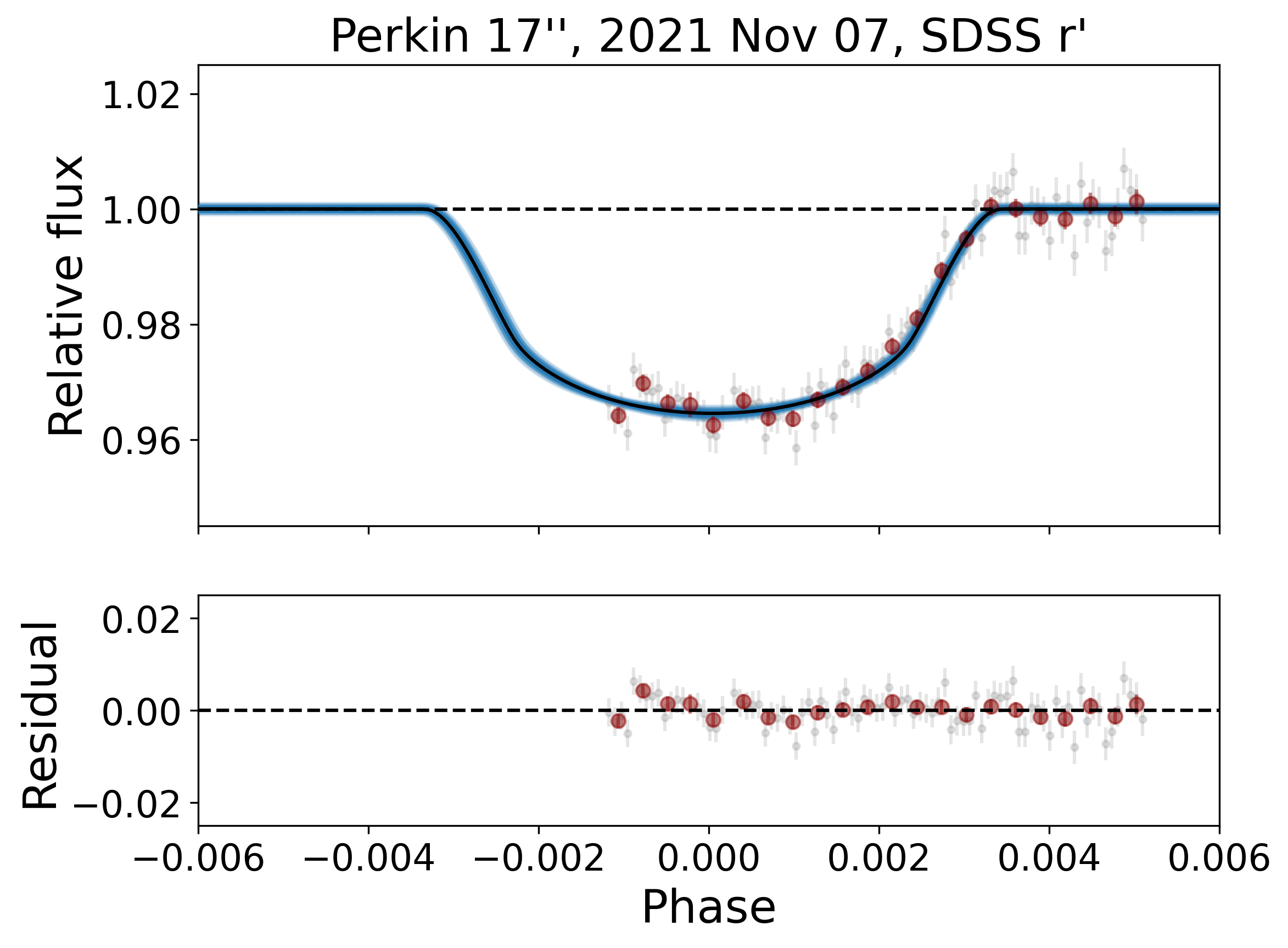}
    \end{minipage}
    \\
    % transits 2
    \begin{minipage}[b]{0.5\linewidth}
    \centering
    \includegraphics[width=\textwidth]{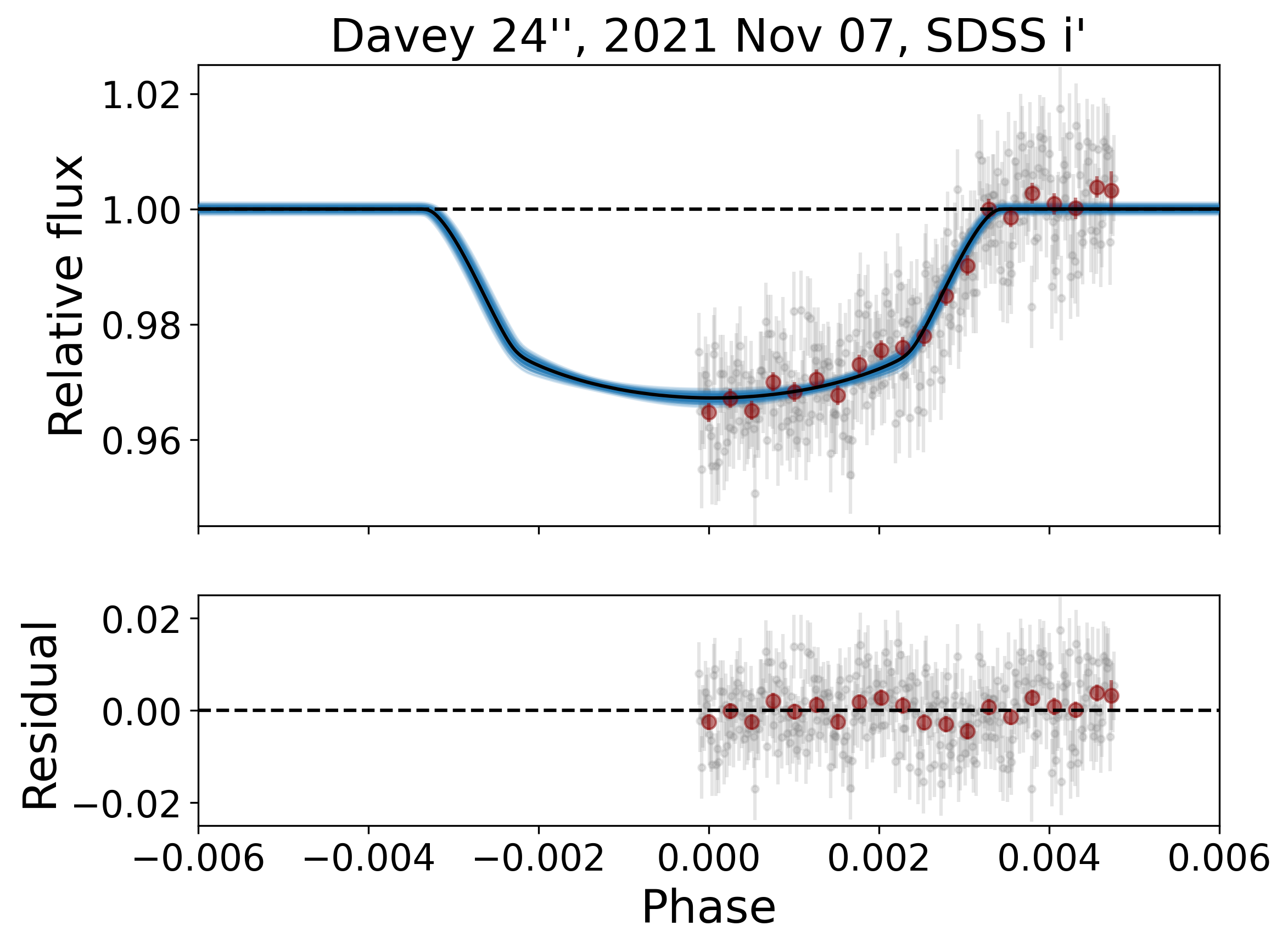}
    \end{minipage}
    \begin{minipage}[b]{0.5\linewidth}
    \centering
    \includegraphics[width=\textwidth]{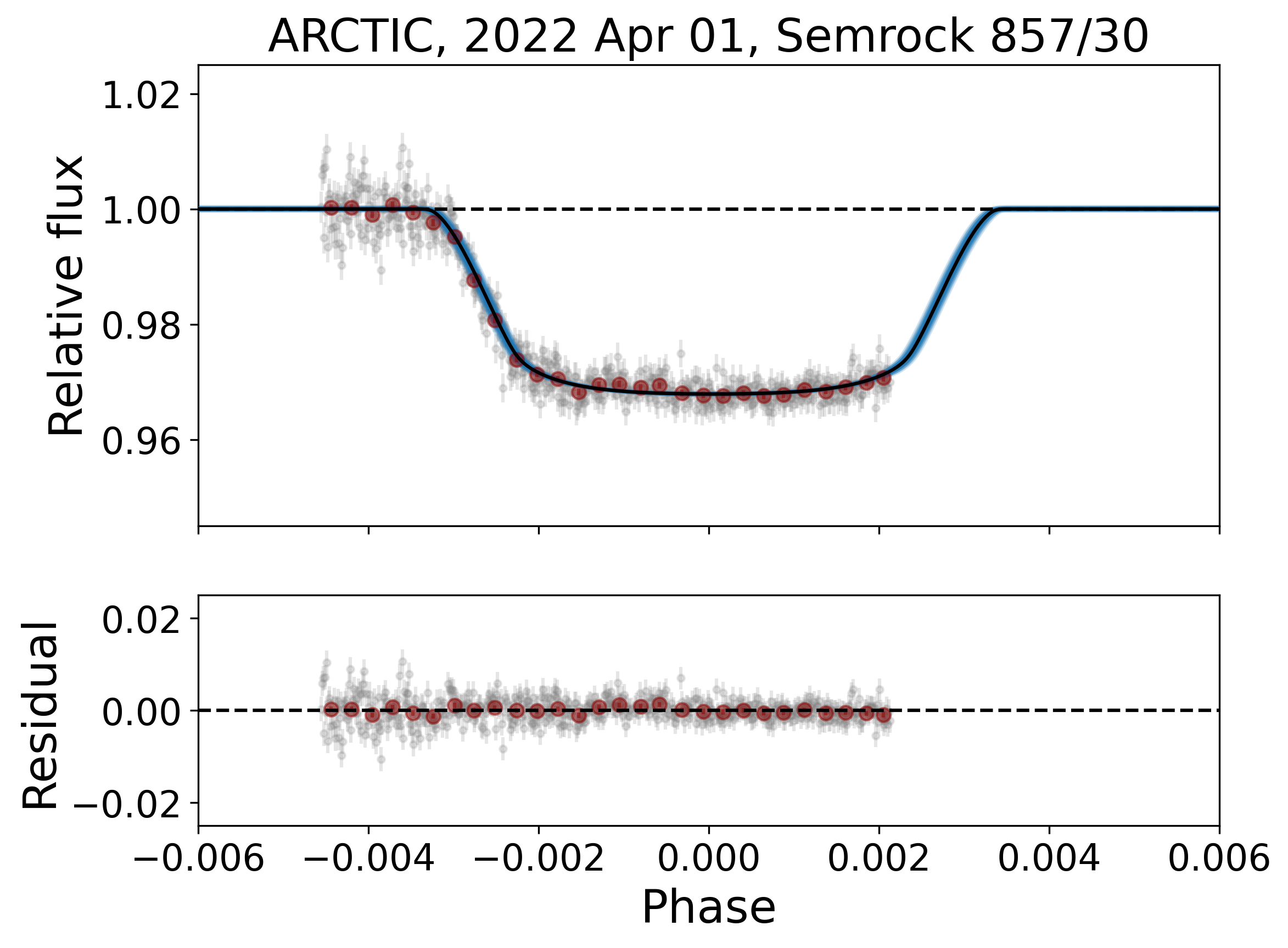}
    \end{minipage}
    \caption{Joint-fit photometry of TOI-1899~b. In each panel, the best-fit model is plotted as a black line, while the blue shaded regions denote the $1\sigma$ (darkest), $2\sigma$, and $3\sigma$ ranges of the derived posterior solution. We show transits from TESS (all available Sectors, phase-folded) (\textit{top left}), the Perkin 17'' in SDSS $r^\prime$ (\textit{top right}), the Davey 24'' in SDSS $i^\prime$ (\textit{bottom left}), and ARCTIC in Semrock 857/30 (\textit{bottom right}). Grey points are unbinned, while dark red points have been binned to a cadence of 10 min. The photometry in this figure is available as Data Behind Figure.}
    \label{fig:joint_fit_phot}
\end{figure*}

\subsection{TESS Photometry}
\label{sec:tess_phot}

TESS first observed TOI-1899 during Sectors 14 (2019 Jul 18 to 2019 Aug 14) and 15 (2019 Aug 15 to 2019 Sep 10) at 2-minute cadence, revealing a single transit event which was used by \citet{Canas2020_toi1899} to derive the initial orbit of this planet. Since then, TESS has re-observed TOI-1899 during Sectors 41 (2021 Jul 23 to 2021 Aug 20), 54 (2022 Jul 09 to 2022 Aug 04), and 55 (2022 Aug 05 to 2022 Sep 01), all at 2-minute cadence. Three more transits were detected during these sectors, allowing us to constrain the planet period much more precisely. We show the TESS transits phase-folded in the {top left panel of \autoref{fig:joint_fit_phot}}. In our analysis, we use the Pre-search Data-Conditioned Simple Aperture Photometry \citep[PDCSAP;][]{Jenkins2016_TESSdata} light curves available from MAST\footnote{Collected in \dataset[DOI:10.17909/SBX7-VG73]{doi.org/10.17909/SBX7-VG73}}, which have already been corrected for dilution. However, in our joint fit, we let the TESS dilution float---see Section \ref{sec:joint_fit} for details.

\subsection{Ground-based Photometry}
\label{sec:ground_phot}

In our efforts to better characterize the orbit of TOI-1899~b, we made several attempts at observing a ground-based transit of the planet, using the single-transit ephemeris from \citet{Canas2020_toi1899}. We monitored TOI-1899 with the 0.6m telescope at Red Buttes Observatory \citep[RBO;][]{Kasper2016_rbo} every night from 2021 June 11 through June 17, but did not see a transit on any of these nights\footnote{Our updated ephemeris reveals the transit midpoint to have been 2021 June 15 at UT 12:52 $\pm$ 0:03---unfortunately not visible from RBO, with the transit starting during morning twilight.}. It was not until a second TESS transit was detected in Sector 41 that we were able to constrain the planet period well enough for successful ground-based transit photometry; we detail the three successful observations below.

\subsubsection{Perkin Observatory 17'' Telescope}

We observed a partial transit of TOI-1899~b on the night of 2021 November 07 with the 17'' (0.43m) PlaneWave Corrected Dall-Kirkham (CDK) telescope at the Richard S. Perkin Observatory at Hobart \& William Smith Colleges. The processed data is shown in the top right panel of \autoref{fig:joint_fit_phot}; due to the long transit duration, we were only able to observe mid-transit through egress.

The Perkin 17'' is mounted on a Paramount equatorial mount and equipped with an SBIG 8300~M camera at Cassegrain focus. The detector array is $3326 \times 2504$ pixels, with a FOV of $\sim21\arcmin \times 16\arcmin$, resulting in an unbinned pixel scale of 0.38\arcsec/pix. We carried out our observations defocused to $\sim$4\arcsec~FWHM in SDSS $r^\prime$, in \mbox{$1 \times 1$} binning, with an exposure time of 180~s. The target started at an airmass of 1.02 and set to 1.85 over the course of our observations. We performed aperture photometry in \texttt{AstroImageJ} \citep[\texttt{AIJ};][]{Collins2017_aij}, following the methodology outlined in \citet{Stefansson2017_diffuser,Stefansson2018_k2_diffuser}. For our final \texttt{AIJ} reduction, we used an object aperture radius of 12 pixels (4.6\arcsec), and inner and outer sky radii of 30 and 45 pixels (11.4\arcsec~and 17.1\arcsec), respectively. 

\subsubsection{Davey Lab 24'' Telescope}

On the same night (2021 November 07), we also observed TOI-1899~b with the 24'' (0.61m) PlaneWave CDK located on the roof of Penn State's Davey Laboratory (\autoref{fig:joint_fit_phot}, bottom left). Similar to the Perkin observations above, we were only able to observe mid-transit through egress.

The Davey Lab 24'' has an SBIG \mbox{STX-9000} camera with an array of $3056 \times 3056$ pixels and a FOV of $\sim32\arcmin \times 32\arcmin$, corresponding to an unbinned pixel scale of 0.63\arcsec/pix. We carried out our observations defocused to $\sim$3\arcsec~FWHM in SDSS $i^\prime$, in $2 \times 2$ binning, with an exposure time of 30~s. During our observations, the target set from an airmass of 1.05 to 1.76. We reduced this photometry with \texttt{AIJ} in the same way as described above, with an object aperture radius of 6 binned pixels (7.5\arcsec), and inner and outer sky radii of 13 and 18 pixels (16.2\arcsec~and 22.5\arcsec). 

\subsubsection{ARCTIC}

We also observed a partial transit of TOI-1899~b on the night of 2022 April 01\footnote{We also attempted observations on this night from the Perkin 17'', the Davey Lab 24'', and the Peter van de Kamp Observatory at Swarthmore College, but were unsuccessful due to weather.} with the Astrophysical Research Consortium (ARC) Telescope Imaging Camera \citep[ARCTIC;][]{Huehnerhoff2016_ARCTIC} on the ARC 3.5m Telescope at Apache Point Observatory ({\autoref{fig:joint_fit_phot}, bottom right}). This time, we were able to observe ingress through mid-transit, but unable to catch egress due to morning twilight.

We operated ARCTIC in the quad-amplifier and fast-readout modes, defocused to $\sim$5\arcsec~FWHM and binned $4 \times 4$ with an exposure time of 25~s. Since our observations began at very high airmass, we used a Semrock 857/30 filter---this narrow passband (842--872~nm) was chosen to avoid atmospheric absorption bands, minimizing the impact of clouds or high airmass on the photometry \citep{Stefansson2017_diffuser, Stefansson2018_diffuserSPIE}. Over the course of the night, the target rose from its starting airmass of 6.11 to end at 1.10. For the \texttt{AIJ} reduction, we used an object aperture radius of 12 binned pixels (5.4\arcsec), and inner and outer sky radii of 20 and 25 pixels (9.1\arcsec~and 11.4\arcsec). We see a strong slope in the raw data from this night; without baseline available on both sides of the transit, we choose to detrend with a line and allow the dilution of this transit to float in our joint fit.

\subsection{Archival Photometry}
\label{sec:archival_phot}

TOI-1899~b is a long-period planet in a crowded field, exemplifying a class of TESS planet candidates which can be difficult to validate. The short TESS sector lengths mean large uncertainties on the period---especially if only a single transit is detected, as was true when \citet{Canas2020_toi1899} initially validated TOI-1899~b---and the large size of TESS pixels on-sky leads to uncertainty about the true host star of the transit and the level of background contamination. Typically, these parameters are constrained through a combination of ground-based transits, which are difficult to schedule without a precise period, and RVs, which are necessarily time-intensive for long-period planets. However, it is possible to glean some of this valuable information from the archival data of long-running photometric surveys, especially for deep transits like those of gas giants around M-dwarfs.

\begin{figure*}[!p]
    \centering
    \includegraphics[width=0.9\textwidth]{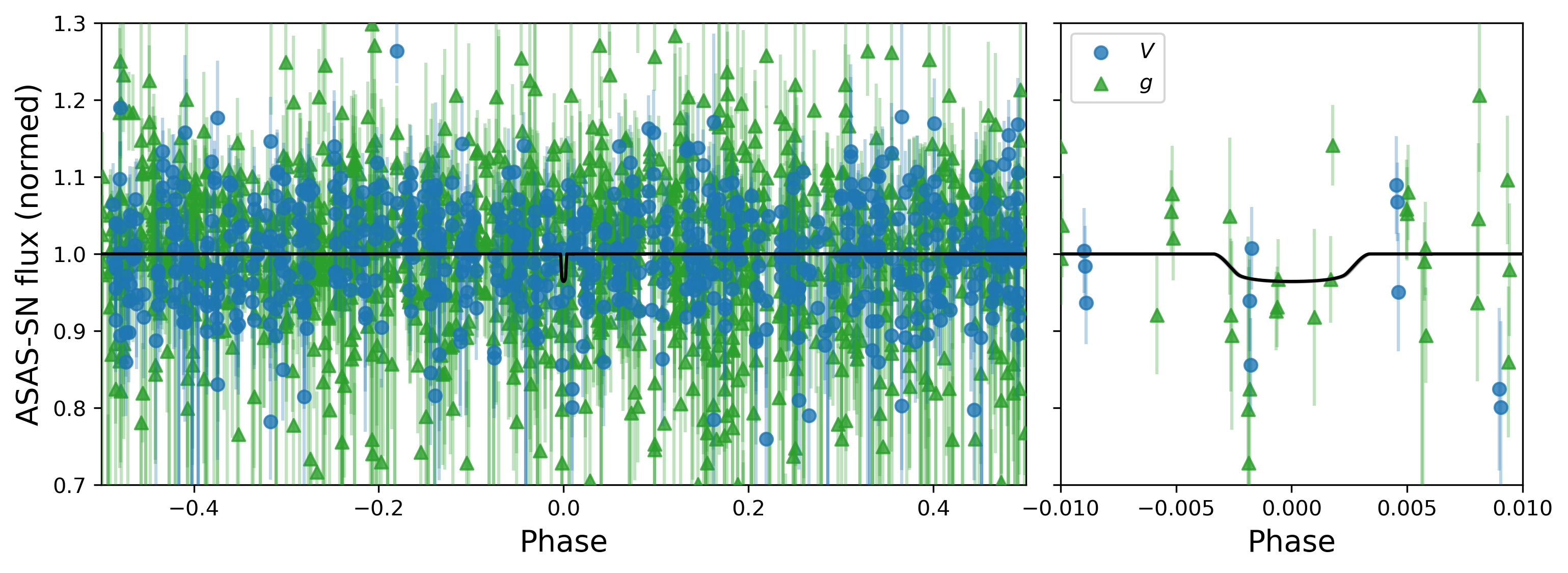}
    \includegraphics[width=0.9\textwidth]{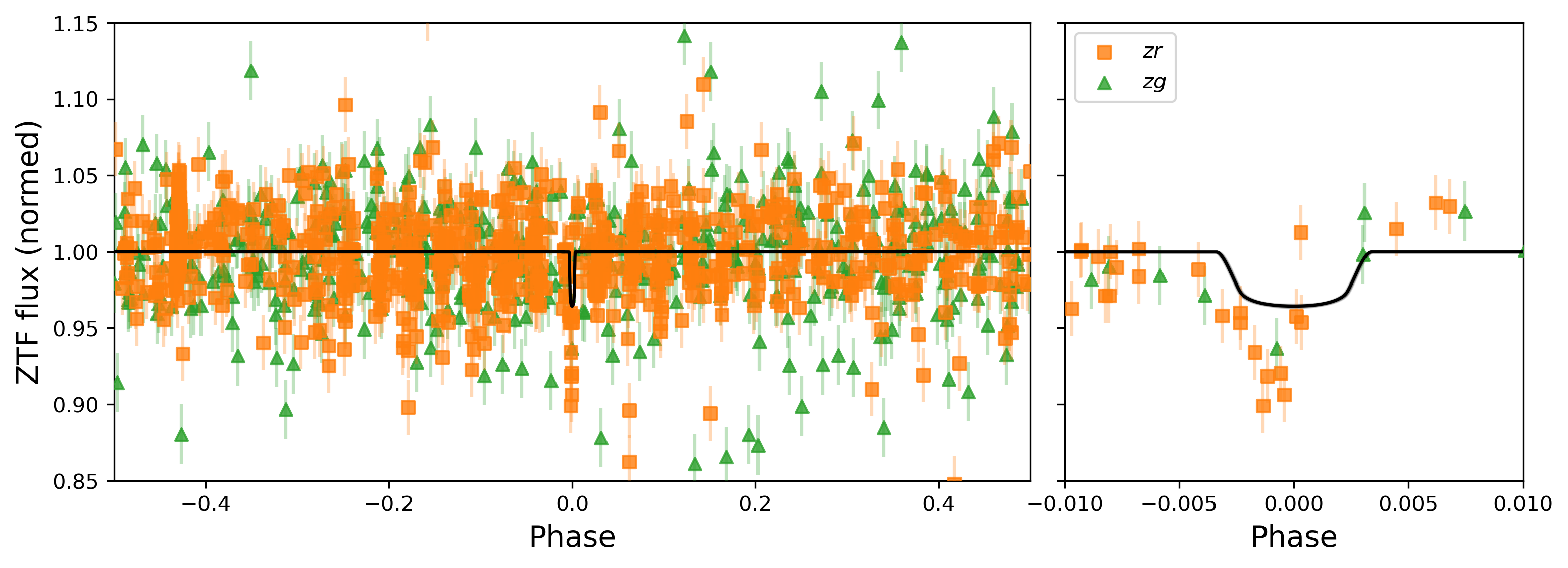}
    \includegraphics[width=0.9\textwidth]{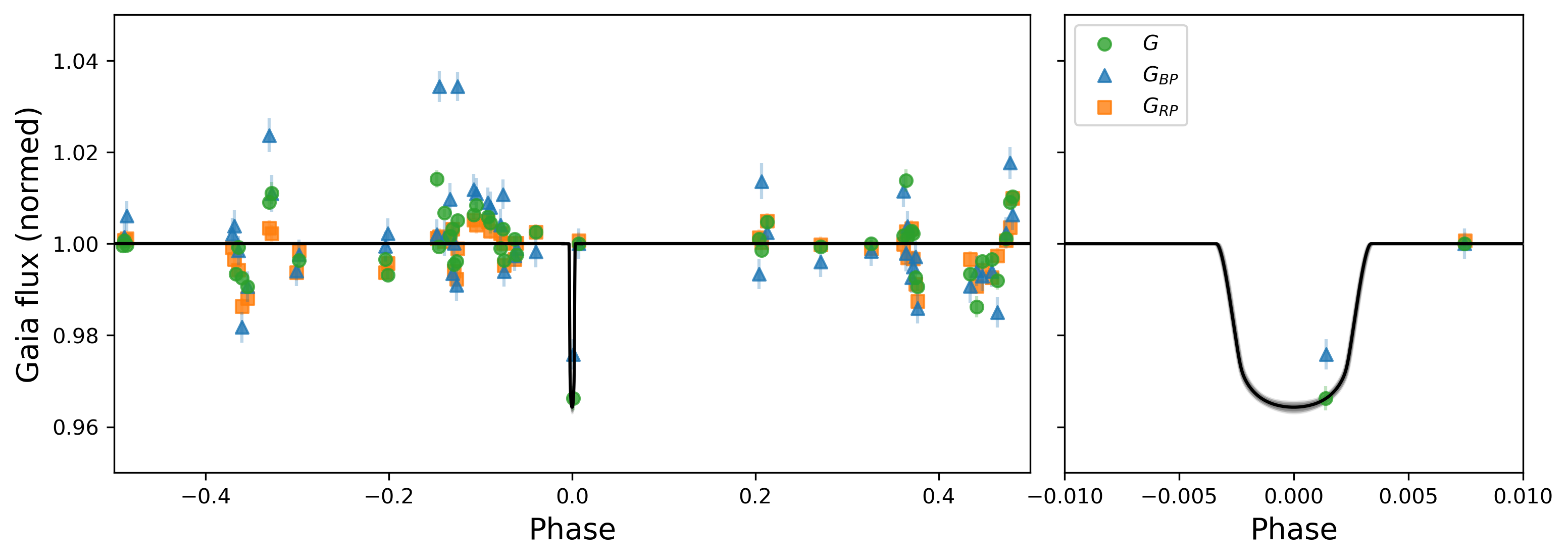}
    \caption{ASAS-SN, ZTF, and \textit{Gaia} photometry phased to the ephemeris of TOI-1899~b, with the right panels zoomed in around the transit. The joint fit model from \autoref{fig:joint_fit_phot} is overplotted, including the 1, 2, and $3\sigma$ posteriors. ASAS-SN does not have the precision to detect the transit, but we see a potential transit in ZTF and a more definitive one in \textit{Gaia}.}
    \label{fig:epochphot}
\end{figure*}

We thus investigate whether the transit of TOI-1899~b can be detected in photometry from the All-Sky Automated Survey for SuperNovae \citep[ASAS-SN;][]{Kochanek2017_asassn} and the Zwicky Transient Facility \citep[ZTF;][]{Masci2019_ztf}. With a $\sim$29 day period, there is only an 0.7\% chance that TOI-1899~b is in-transit at any given time, but these surveys' years of coverage encompass hundreds of observations. We retrieve the publicly accessible data from ASAS-SN Sky Patrol in $V$ and $g$ (2015 Feb 24 to 2018 Nov 10, and 2018 Apr 12 to 2022 Dec 22, respectively) and ZTF DR17 in $zr$ and $zg$ (including data up through 2023 Mar 9), and phase them to the planetary ephemeris as derived in this paper. \autoref{fig:epochphot} shows that unfortunately, ASAS-SN does not have the photometric precision necessary to detect the transit, but we do see a dip in flux suggestive of a transit in ZTF, especially in the $zr$ band.

In addition, TOI-1899 is among the subset of \textit{Gaia} sources with epoch photometry available in DR3 \citep{gaia_dr3}. The idea of using epoch photometry from astrometric missions such as \textit{Gaia} or \textit{Hipparcos} to detect transiting exoplanets is not new---e.g., the recovery of HD 209458~b from \textit{Hipparcos} data \citep{robichon_2000_hipparcos_transit}, or the recent discoveries of \textit{Gaia}-1~b and \textit{Gaia}-2~b \citep{panahi_2022_gaia_transit}---but the excellent spatial resolution and photometric precision of \textit{Gaia} also make it useful for validating TESS objects like TOI-1899~b. However, the cadence of \textit{Gaia} data is much sparser, so one must be fortunate enough for an observation epoch to line up with the transit.

We retrieve the epoch photometry using the \textit{Gaia} archive DataLink service\footnote{Directly accessible at \url{https://gea.esac.esa.int/data-server/data?retrieval_type=epoch_photometry&ID=Gaia+DR3+2073530190996615424&format=csv}}, discard any data which have been rejected by either the photometry or variability processing pipelines, and normalize the $G$, $G_{\rm{RP}}$, and $G_{\rm{BP}}$ bands separately to their median fluxes. The bottom panel of \autoref{fig:epochphot} shows that the most significant drop in flux clearly lines up with the transit. If we only consider the $G_{\rm{BP}}$ photometry, we might dismiss this dip as random scatter, as we see similar excursions in flux at other orbital phases. However, the concurrent $G$ photometry shows the same dip while exhibiting far less scatter overall, suggesting that this flux decrease is real and produced by the planetary transit.

While we do not include these datasets in our subsequent analysis due to their large scatter and/or sparse cadence, we present this exercise to demonstrate that long-term photometric surveys and \textit{Gaia} epoch photometry---when released for all sources in the upcoming DR4---can potentially be used to validate deep TESS transits, such as those of giant planet candidates around M-dwarfs, and help refine ephemerides as a result of their longer time baselines.

\section{Stellar Parameters} % DONE
\label{sec:stellar}

Initial stellar parameters for the host star TOI-1899 were derived by \citet{Canas2020_toi1899} using spectroscopic parameters from \texttt{HPF-SpecMatch} \citep{Stefansson2020_g940} as priors for an SED fit with the \texttt{EXOFASTv2} analysis package \citep{eastman_2019_exofastv2}. Since then, the \texttt{HPF-SpecMatch} spectral library has been expanded from 55 stars to 166, now spanning $T_{\rm{eff}}$ = 2700 to 6000~K, $\log{g}$ = 4.29 to 5.26, and [Fe/H] = \mbox{-0.5} to 0.5. Thus, we repeat the procedure outlined in Section 4.3 of \citet{Canas2020_toi1899} to re-derive the stellar parameters.

\texttt{HPF-SpecMatch} is broadly based on the algorithms for \texttt{SpecMatch-Emp} outlined by \citet{Yee2017_specmatch}. In brief, it compares a high-S/N HPF spectrum of the target star to a library containing similarly high S/N spectra (also observed with HPF) of stars with well-characterized properties. Using $\chi^2$ minimization, it produces a weighted linear combination of the five best-fitting library spectra, from which it determines the target's spectroscopic parameters. The errors are determined by a leave-one-out cross-validation, which estimates the properties of a library star of interest using all the other stars in the library; we take the difference between these derived values and the true values as the errors. 

\begin{figure*}[!htbp]
    \begin{minipage}[b]{0.6\linewidth}
    \centering
    \includegraphics[width=\textwidth]{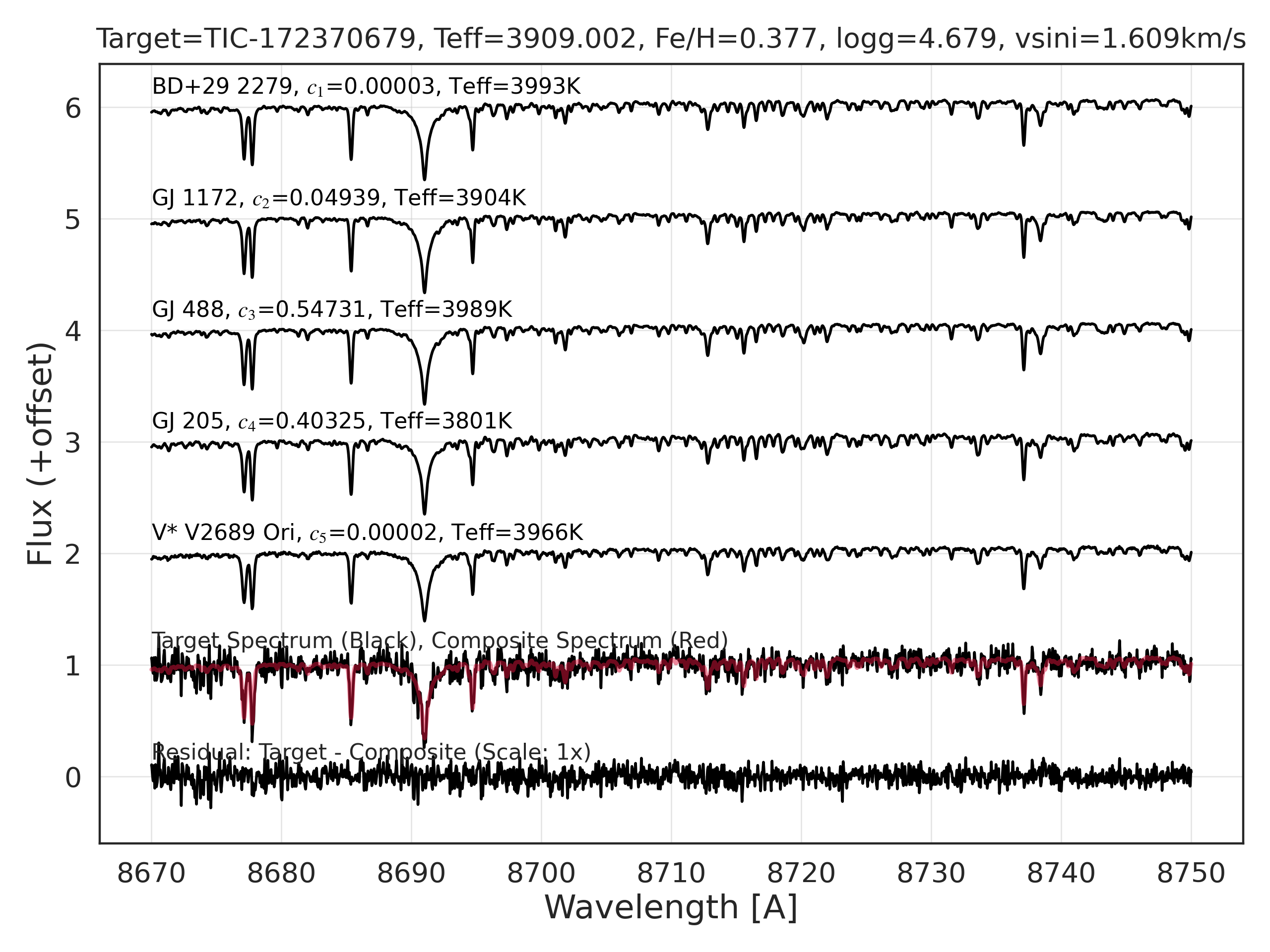}
    \end{minipage}
    \begin{minipage}[b]{0.4\linewidth}
    \centering
    \includegraphics[width=\textwidth]{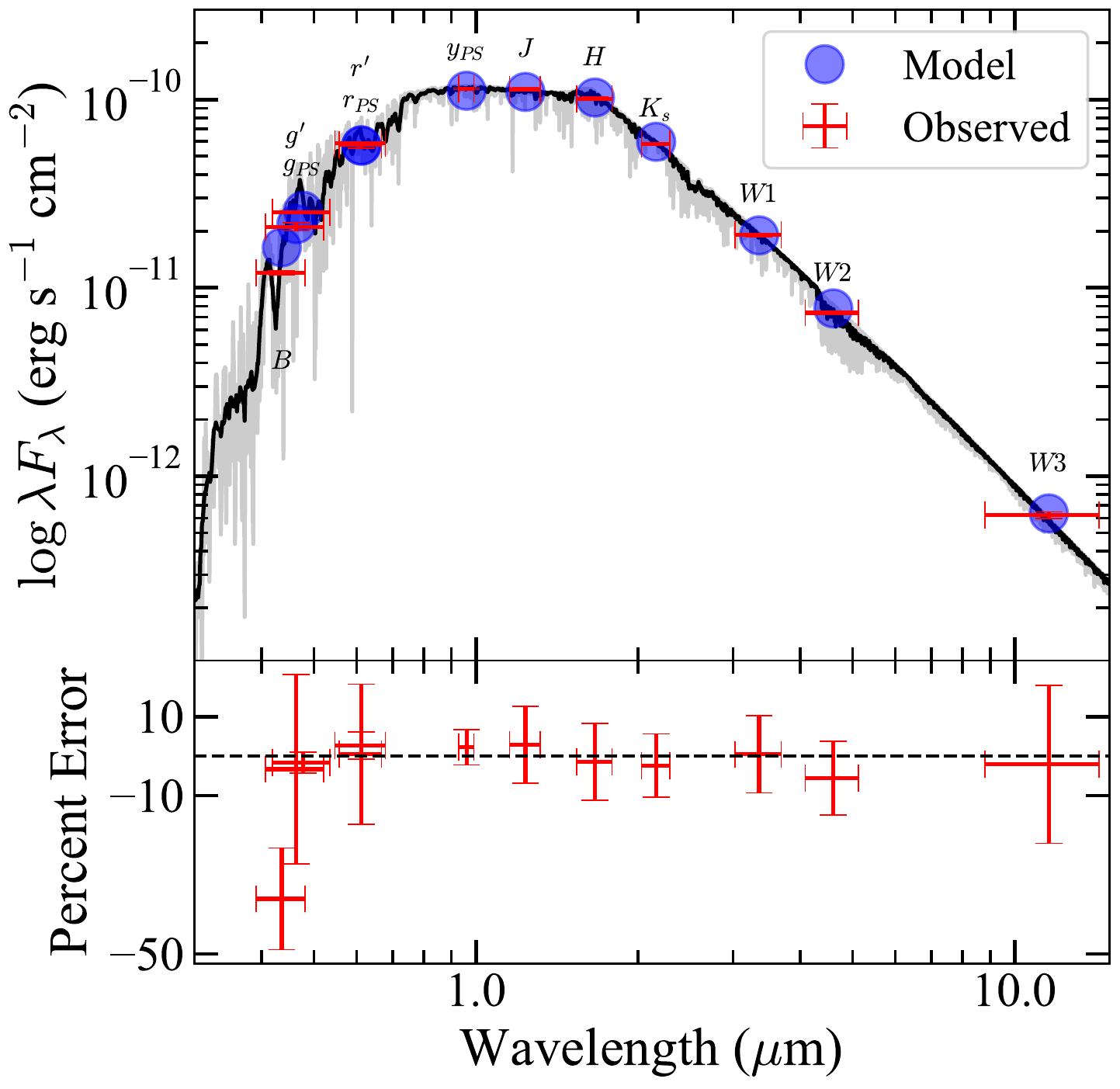}
    \end{minipage}
    \caption{\textit{Left}: Spectra of the five best-matching \texttt{HPF-SpecMatch} library stars, with the composite library spectrum (\textit{red}) overplotted against the highest-S/N observed spectrum of the target star TOI-1899. The composite spectrum is constructed using the weights $c_i$ noted for each library star. \textit{Right}: SED of TOI-1899 as fit by \texttt{EXOFASTv2}, showing the broadband photometric measurements (\textit{red}; $x$-errorbars indicate the passband width) and the derived MIST model fluxes (\textit{blue}). A NextGen BT-SETTL model \citep{allard_2012_btsettl} is overlaid for reference (\textit{grey}, smoothed version in black), and is not used as part of the SED fit.}
    \label{fig:specmatch_sed}
\end{figure*}

We use the single highest-S/N HPF spectrum of TOI-1899 to perform this comparison against the updated \texttt{HPF-SpecMatch} spectral library---taken on 2019 Nov 23, this happens to be the exact same spectrum used by \citet{Canas2020_toi1899} when they derived spectroscopic parameters from the old \texttt{HPF-SpecMatch} library. In the left panel of \autoref{fig:specmatch_sed}, we show the spectra of the five best-fitting library stars across a portion of HPF order index 5, and the composite spectrum compared to that of TOI-1899.

We then repeat the subsequent SED fit (\autoref{fig:specmatch_sed}, right) with \texttt{EXOFASTv2}, using the default MIST stellar models \citep{Choi2016_mist, Dotter2016_mist} and the priors listed in \autoref{tab:stellar_params}. We use the same priors as \citet{Canas2020_toi1899}, apart from the new $T_{\rm{eff}}$, [Fe/H], and $\log{g}$ derived from the updated \texttt{HPF-SpecMatch} library and the addition of the Pan-STARRS DR2 $gry$ photometry. We find that all derived stellar parameters are consistent with the values reported in Table 2 of \citet{Canas2020_toi1899}. Nevertheless, we present the updated stellar parameters in \autoref{tab:stellar_params} for completeness and adopt these new values for use in our joint fit.

% \startlongtable
\begin{deluxetable*}{llcccrr}[!hbp]
\tablecaption{Summary of Updated Stellar Parameters}
\tablehead{\colhead{~~~~Parameter\tablenotemark{a}} & \colhead{Units} & \colhead{Source} & \colhead{\citet{Canas2020_toi1899}} & \colhead{This Work} & \colhead{\% Diff.\tablenotemark{b}} & \colhead{$\sigma$ Diff.\tablenotemark{b}}}
\tablecolumns{7}
\startdata
\sidehead{Priors for \texttt{EXOFASTv2} SED fit:}
\sidehead{~~~~\textit{Spectroscopic Parameters:}}
~~~~Effective Temperature & $T_{\rm eff}$ (K) & \texttt{HPF-SpecMatch} & 
$3925\pm77$ & $3909\pm88$ & --4.1 & --0.21 \\
~~~~Metallicity & $[{\rm Fe/H}]$ (dex) & \texttt{HPF-SpecMatch} & 
$0.20\pm0.13$ & $0.38\pm0.12$ & +90.0 & +1.38 \\
~~~~Surface gravity & $\log{g}$ (cgs) & \texttt{HPF-SpecMatch} & 
$4.68\pm0.05$ & $4.68\pm0.05$ & $<$ 0.1 & $<$ 0.01 \\
\sidehead{~~~~\textit{Photometric Magnitudes:}}
~~~~APASS Johnson $B$ & $B$ (mag) & \citet{henden_2015_apass} & 
$15.898 \pm 0.029$ & same & --- & --- \\
~~~~APASS Sloan $g^\prime$ & $g^\prime$ (mag) & \citet{henden_2015_apass} &
$15.115 \pm 0.054$ & same & --- & --- \\
~~~~APASS Sloan $r^\prime$ & $r^\prime$ (mag) & \citet{henden_2015_apass} &
$13.728 \pm 0.040$ & same & --- & --- \\
~~~~Pan-STARRS $g_{\rm{PS}}$ & $g_{\rm{PS}}$ (mag) & \citet{chambers_2016_panstarrs} &
--- & $14.889 \pm 0.006$ & --- & --- \\
~~~~Pan-STARRS $r_{\rm{PS}}$ & $r_{\rm{PS}}$ (mag) & \citet{chambers_2016_panstarrs} &
--- & $13.708 \pm 0.008$ & --- & --- \\
~~~~Pan-STARRS $y_{\rm{PS}}$ & $y_{\rm{PS}}$ (mag) & \citet{chambers_2016_panstarrs} &
--- & $12.495 \pm 0.010$ & --- & --- \\
~~~~2MASS $J$ & $J$ (mag) & \citet{cutri_2003_2mass} &
$11.342 \pm 0.022$ & same & --- & --- \\
~~~~2MASS $H$ & $H$ (mag) & \citet{cutri_2003_2mass} &
$10.666 \pm 0.022$ & same & --- & --- \\
~~~~2MASS $K_s$ & $K_s$ (mag) & \citet{cutri_2003_2mass} &
$10.509 \pm 0.018$ & same & --- & --- \\
~~~~WISE $W1$ & $W1$ (mag) & \citet{wright_2010_wise} &
$10.412 \pm 0.022$ & same & --- & --- \\
~~~~WISE $W2$ & $W2$ (mag) & \citet{wright_2010_wise} &
$10.460 \pm 0.021$ & same & --- & --- \\
~~~~WISE $W3$ & $W3$ (mag) & \citet{wright_2010_wise} &
$10.312 \pm 0.045$ & same & --- & --- \\
\sidehead{~~~~\textit{Other Priors:}}
~~~~Distance & $d$ (pc) & \citet{bailer-jones_2018_distance} & 
$128.4 \pm 0.3$ & same & --- & --- \\
~~~~Maximum $V$ extinction & $A_V$ (mag) & \citet{green_2019_extinction} & 
$0.02$ & same & --- & --- \\
\sidehead{Derived Stellar Parameters (used for joint RV + transit fit):}
~~~~Effective Temperature & $T_{\rm eff}$ (K) & \texttt{EXOFASTv2} &
$3841^{+54}_{-45}$ & $3926^{+45}_{-47}$ & +2.2 & +1.57 \\
~~~~Metallicity & $[{\rm Fe/H}]$ (dex) & \texttt{EXOFASTv2} &
$0.31^{+0.11}_{-0.12}$ & $0.28\pm0.11$ & --9.6 & --0.25 \\
~~~~Surface gravity & $\log{g}$ (cgs) & \texttt{EXOFASTv2} &
$4.669^{+0.025}_{-0.022}$ & $4.672^{+0.021}_{-0.020}$ & +0.1 & +0.12 \\
~~~~Mass & $M_*$ (\msun) & \texttt{EXOFASTv2} &
$0.627^{+0.026}_{-0.028}$ & $0.632^{+0.026}_{-0.025}$ & +0.7 & +0.19 \\
~~~~Radius & $R_*$ (\rsun) & \texttt{EXOFASTv2} &
$0.607^{+0.019}_{-0.023}$ & $0.607^{+0.017}_{-0.016}$ & $<$ 0.1 & $<$ 0.01 \\
~~~~Density & $\rho_*$ (cgs) & \texttt{EXOFASTv2} &
$3.95^{+0.37}_{-0.29}$ & $3.98^{+0.29}_{-0.27}$ & +0.7 & +0.08 \\
~~~~Age & Age (Gyr) & \texttt{EXOFASTv2} &
$7.4^{+4.4}_{-4.6}$ & $7.1^{+4.5}_{-4.8}$ & --4.1 & --0.07 \\
~~~~V-band extinction & $A_V$ (mag) & \texttt{EXOFASTv2} &
$0.010\pm0.007$ & $0.010\pm0.007$ & $<$ 0.1 & $<$ 0.01 \\
\enddata
\label{tab:stellar_params}
\tablenotetext{a}{See Table 3 in \citet{Eastman2017_exofastv2} for a detailed description of all parameters}
\tablenotetext{b}{Values from \citet{Canas2020_toi1899} used as the fiducial}
\end{deluxetable*}

\subsection{Stellar Activity}
\label{sec:activity}

With a total of 44 HPF RVs now spanning nearly 3 years, we investigated whether these RVs were correlated with several common NIR activity indicators---the differential line width (dLW), the chromatic index \citep[CRX;][]{Zechmeister2018_serval}, and the Ca infrared triplet (Ca IRT; air wavelengths of 8498, 8542, and 8662 $\text{\AA}$) indices. We found no strong correlations between the RVs and any of these activity metrics using the Kendall rank correlation coefficient (Kendall $\tau$). We include the generalized Lomb-Scargle (GLS) periodograms and corner plot for the HPF RVs and activity indicators in \autoref{app:hpf_activity}, with the activity index values available as Data Behind Figure. With only 4 NEID RVs, we do not have enough data to look for similar temporal correlations, but examination of the raw spectra shows no signs of emission in either the H$\alpha$ line or the Na doublet\footnote{While NEID's wavelength range also includes the Ca II H\&K lines, we did not use them because these orders are very low S/N for a faint M-dwarf like TOI-1899.}.

We note that in both HPF and NEID spectra, all three lines of the Ca IRT appear to show consistent low levels of emission in the line core, but this does not appear to have any significant effect on the RVs, as we see nearly no correlation between any of the Ca IRT indices and the HPF RVs ($\tau$ = 0.0023 -- 0.0035). Given that none of the other indicators we examined show signs of activity, we conclude that TOI-1899 is a quiet, low-activity M-dwarf.

\begin{figure*}[!htbp]
    \centering
    \includegraphics[width=0.9\textwidth]{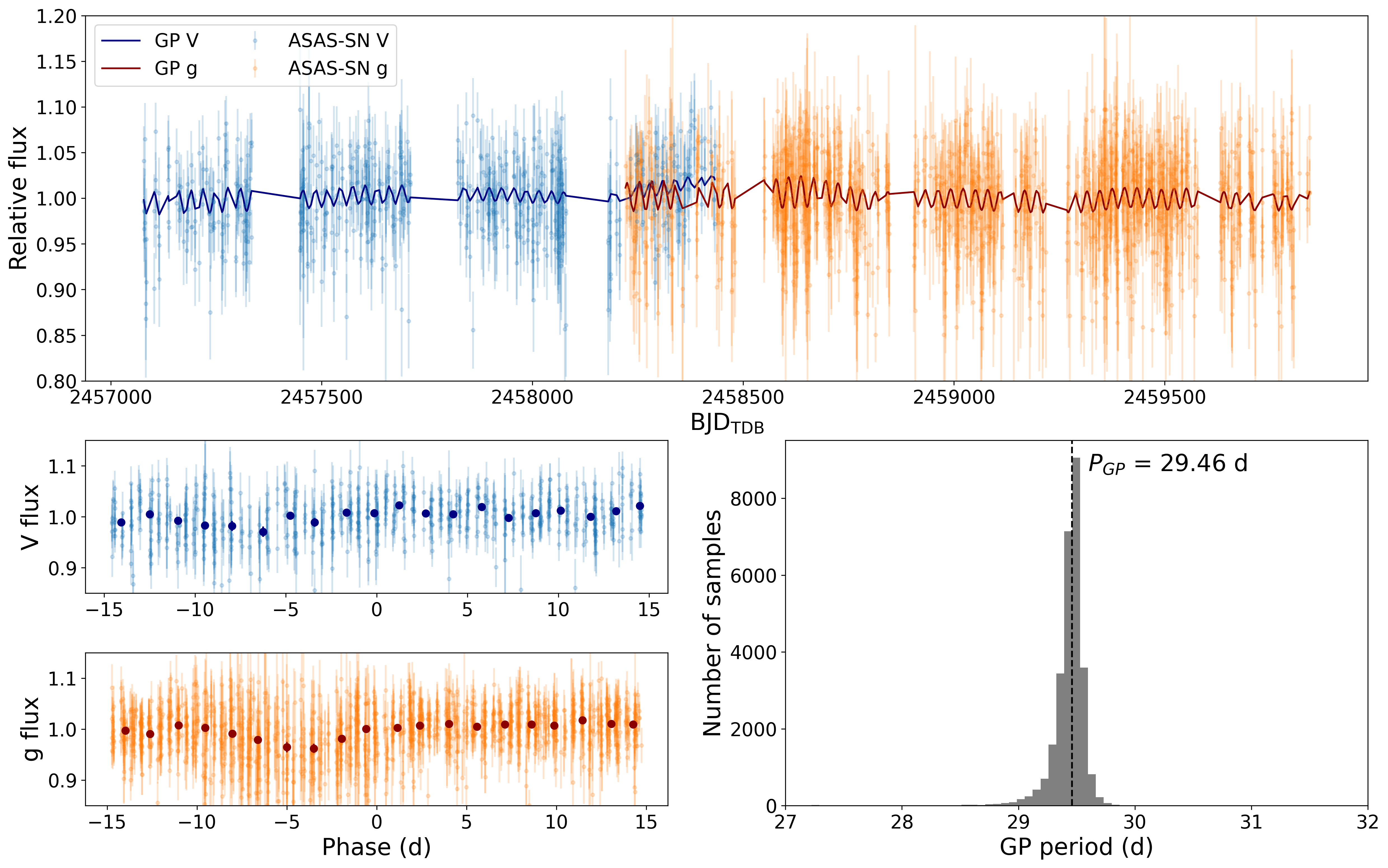}
    \caption{\textit{Top}: ASAS-SN photometry fit with a \texttt{celerite} quasi-periodic kernel, suggesting a stellar rotation period of $29.46^{+0.07}_{-0.10}$ days. \textit{Bottom left}: Photometry phased to the derived GP period. Dark points have been binned to better show the flux variation as a function of phase. \textit{Bottom right}: Histogram of posterior samples for $P_{GP}$.}
    \label{fig:prot_gp}
\end{figure*}

\subsection{Stellar Rotation Period}
\label{sec:rotation}

In an effort to constrain the stellar rotation period, we examine the GLS periodograms of the previously-mentioned HPF activity indicators (dLW, CRX, and Ca IRT). We find no signals with False Alarm Probability (FAP) $<$~0.1\%. We do see peaks at the FAP $\sim$1\% level in the Ca IRT 3 (8662 $\text{\AA}$) index, with periodicities $\sim$80 and $\sim$100~d, but it is doubtful whether these peaks correspond to rotationally-modulated activity, as they are not corroborated by any other activity indicator---including the other two Ca IRT indices.

We also search for rotationally-modulated photometric variability in the TESS photometry, as well as the publicly accessible photometry from ASAS-SN and ZTF (as mentioned in Section \ref{sec:archival_phot}). Since the ASAS-SN $V$ and $g$ photometry share only a small time baseline overlap and the bandpasses are similar, we also try combining these datasets to see if this boosts the significance of long-period signals. We include the GLS periodograms for ASAS-SN ($V$, $g$, and combined), ZTF ($zr$ and $zg$), and TESS in \autoref{app:prot}.

From the combined ASAS-SN $V+g$ periodogram (\autoref{fig:pgram_asassn}), we detect multiple peaks around $P \sim$ 27--29 days with FAP $< 0.1\%$. A similar period ($P \sim$ 31~d) is suggested by ZTF $zr$ (\autoref{fig:pgram_ztf}), albeit with considerable aliasing as a result of the window function. Unfortunately, these peaks cannot be corroborated by the TESS PDCSAP photometry, which is unreliable when used to detect photometric variability on periods $\gtrsim$ 10~d, as a result of co-trending performed by the PDC processing \citep[as discussed by e.g.,][]{claytor_2022_tess_rot}.

However, the TESS light curves can still be used to search for signs of shorter-period variability. We find many broad peaks in the $P \sim$ 4--15~d regime that are all formally highly significant (FAP $\ll 0.1\%$), but we also observe that these periods change from sector to sector, including back-to-back sectors (most strikingly, Sectors 54 and 55 in \autoref{fig:pgram_tess}). Therefore, we find it unlikely that the observed photometric variability in TESS PDCSAP is directly tracing the stellar rotation period, which should remain constant. In addition, neither ASAS-SN nor ZTF find any strong evidence of periodicity in the 4--15~d period range, even though these datasets should have the requisite sampling to detect them. Furthermore, we do not see these peaks when applying the systematics-insensitive periodogram from \texttt{TESS-SIP} \citep{hedges_2020_tess-sip}, which attempts to detrend out the effects of TESS instrumental systematics such as scattered light. Using either the target pixel files (TPFs) or light curve files (LCFs), we find that the strongest peaks in the \texttt{TESS-SIP} periodogram line up with excess power in background pixels (\autoref{fig:pgram_tess-sip}), suggesting their cause is an instrumental effect not limited to our target star.

Since the ASAS-SN data is the best-behaved and longest-running (spanning $\sim$7 years) of our photometric datasets, we also attempt to fit a Gaussian Process to it using the \texttt{celerite} quasi-periodic kernel \citep{foreman-mackey_fast_2017} with the period shared between the $V$ and $g$ photometry, using a uniform prior from 1--100 days (\autoref{fig:prot_gp}). The quasi-periodic GP fit yields a period of $29.46^{+0.07}_{-0.10}$ days for the photometric modulation---intriguingly close to the orbital period of TOI-1899~b.

This is also very close to the lunar synodic period of 29.53~d. We do not see power in the window function at this period, suggesting the detected period is not an consequence of the observing cadence, though it is possible that lunar contamination could produce periodic photometric variations. We tentatively adopt this as the stellar rotation period, noting that it would be consistent with the lack of visible broadening in either the HPF or NEID spectra, which indicates $v \sin i \lesssim 2$ km/s.

\section{Joint Fitting of Transit \& RV Data} % DONE
\label{sec:joint_fit}

We jointly model the photometry and RVs using \texttt{juliet} \citep{Espinoza2019_juliet}, which relies on \texttt{batman} \citep{Kreidberg2015_batman} for the photometric modeling and \texttt{radvel} \citep{Fulton2018_radvel} for the RV modeling. The photometric transit model is based on that of \citet{Mandel2002_analytic}, and uses the $q_1$ and $q_2$ parameterization from \citet{Kipping2013_quadld} for the quadratic limb-darkening law. \texttt{juliet} uses the dynamic nested sampling package \texttt{dynesty} \citep{Speagle2020_dynesty} to perform its parameter estimation. We model the RVs using a standard Keplerian model, and allow the eccentricity to float. For the joint fitting of the photometry and RVs, we also include a simple white-noise model in the form of a jitter term that is added in quadrature to the measurement errors from each dataset. 

For the transit model, we fix the dilution of the Perkin and Davey transits to 1, and allow dilution to float for TESS and ARCTIC. We observed heavy systematics in the raw ARCTIC data, which we attempted to detrend out---but without having baseline on both sides, we were not confident that this dataset accurately reflected the true transit depth. We also experimented with allowing the TESS dilution and noise properties to float independently for each sector, since TOI-1899 lies in a crowded field where the background stellar contamination will realistically change from sector to sector. However, we found that this introduced too many free parameters, making it very difficult for the joint fit to converge, so in our final \texttt{juliet} fit, we group together all 5 sectors of TESS data as a single instrument with shared limb-darkening, dilution, and noise parameters. In addition, we include a GP using the quasi-periodic kernel from \texttt{celerite} \citep{foreman-mackey_fast_2017} for the TESS data in order to fit for correlated noise.

We present the joint fit to our RVs and transit photometry in \autoref{fig:joint_fit_rv} and \autoref{fig:joint_fit_phot}, respectively, and the final derived planet parameters in \autoref{tab:joint_fit_params}, with the values from \citet{Canas2020_toi1899} shown for comparison. Our results broadly agree with the earlier-derived parameters, with the additional transit observations greatly increasing the precision on the orbital period. 
We note that the RV jitter term, $\sigma_{\rm{HPFpre}}$, is somewhat higher in our fit, possibly since the orbital period is now tightly constrained by the transits, whereas it was free to float in the single-transit solution.

Notably, our fit yields a smaller planetary radius of $0.99\pm{0.03}~R_{J}$, compared to the earlier radius estimate of $1.15_{-0.05}^{+0.04}~R_{J}$, with corresponding increases in density and surface gravity. We believe the previous radius value was inflated due to overcorrection of dilution in TESS Sectors 14 and 15, as there was no ground-based transit data available at the time. We find an overall TESS dilution factor of $D_{\rm{TESS}} = 1.086$. This effect has been seen before in, e.g., \citet{burt_2020_toi824}, who found a 13\% difference in derived planetary radius as a result of properly accounting for the TESS dilution overcorrection. This is supported by our reanalysis of the TESS photometry using the TESS-\textit{Gaia} Light Curve package \citep[\texttt{tglc};][]{Han2023_TGLC}, which uses \textit{Gaia} DR3 source positions and brightnesses to construct a model of the TESS PSF, and uses this to perform PSF photometry on the full-frame images \citep[for details see Sections 3 and 4.1 of][]{Han2023_TGLC}. A photometric fit to the \texttt{tglc}-processed TESS light curve yields a scaled planetary radius of $R_p/R_* = 0.169 \pm 0.004$ (corresponding to $R_p = 0.99 \pm 0.03~R_{J}$), which is fully consistent with the radius derived from ground-based photometry. The new radius brings TOI-1899~b in line with the size expected from Jovian planet models, resolving previous concerns about unaccounted-for inflation mechanisms in this planet \citep{muller_2023_giant_ariel}.

We use \texttt{thejoker} \citep{PriceWhelan2017_thejoker} to put constraints on the presence of long-period, co-planar companions from the HPF RV residuals. We put a 3$\sigma$ upper limit of 0.54~$M_{J}$ ($K$ = 36 m/s) on companion objects within 1~au, and 1.16~$M_{J}$ ($K$ = 49 m/s) within 2~au. TOI-1899~b thus appears to be isolated from other massive planets in its system, if they exist at all. Though we can neither confirm nor rule out the presence of small companion planets like those commonly found alongside FGK WJs, the low eccentricity of TOI-1899~b seems to suggest a similar dynamically-cool formation method.

% \startlongtable
\begin{deluxetable*}{llccrr}[!hp]
\tablecaption{Derived Parameters for the TOI-1899 System\label{tab:joint_fit_params}}
\tablehead{\colhead{~~~~Parameter} & \colhead{Units} &
\colhead{\citet{Canas2020_toi1899}} & \colhead{This Work} & \colhead{\% Diff.\tablenotemark{b}} & \colhead{$\sigma$ Diff.\tablenotemark{b}}}
\tablecolumns{6}
\startdata
\sidehead{Orbital Parameters:}
~~~~Orbital Period & $P$ (days) &
$29.02_{-0.23}^{+0.36}$ & $29.090312_{-0.000035}^{+0.000036}$ & +0.2 & +0.20 \\
% ~~~Time of Periastron & $T_P$ (BJD\textsubscript{TDB}) & $2458705.37_{-2.48}^{+2.28}$ \\
~~~~Eccentricity & $e$ & 
$0.118_{-0.077}^{+0.073}$ & $0.044_{-0.027}^{+0.029}$ & --62.7 & --0.96 \\
~~~~Argument of Periastron & $\omega$ (degrees) & 
$-13_{-28}^{+27}$ & $-53_{-36}^{+42}$ & --- & --1.42 \\
~~~~Semi-amplitude Velocity & $K$ (m/s) &
$59.91_{-6.32}^{+6.41}$ & $59.82_{-3.49}^{+3.52}$ & --0.2 & --0.01 \\
~~~~HPF$_{\rm{pre}}$ RV Offset  & $\gamma_{\rm{HPFpre}}$ (m/s) & 
$16.64_{-5.23}^{+5.39}$ & $24.32_{-2.55}^{+2.49}$ & +46.2 & +1.42 \\
~~~~HPF$_{\rm{pre}}$ RV Jitter & $\sigma_{\rm{HPFpre}}$ (m/s) & 
$0.39_{-0.36}^{+3.84}$ & $7.90^{+4.83}_{-4.94}$ & +1925.6 & +1.96 \\
~~~~HPF$_{\rm{post}}$ RV Offset & $\gamma_{\rm{HPFpost}}$ (m/s) & 
--- & $-50.35_{-14.74}^{+18.22}$ & --- & --- \\
~~~~HPF$_{\rm{post}}$ RV Jitter & $\sigma_{\rm{HPFpost}}$ (m/s) & 
--- & $2.47_{-2.39}^{+25.34}$ & --- & --- \\
~~~~NEID RV Offset  & $\gamma_{\rm{NEID}}$ (m/s) & 
--- & $-1.49_{-10.63}^{+10.80}$ & --- & --- \\
~~~~NEID RV Jitter & $\sigma_{\rm{NEID}}$ (m/s) & 
--- & $18.39_{-6.97}^{+13.51}$ & --- & --- \\
\sidehead{Transit Parameters:}
~~~~Time of Conjunction & $T_C$ (BJD\textsubscript{TDB}) & $2458711.957792_{-0.001179}^{+0.001182}$ & $2458711.959171_{-0.001105}^{+0.001116}$ & --- & +1.16 \\
~~~~Scaled Radius & $R_{p}/R_{*}$  & 
$0.194_{-0.005}^{+0.004}$ & $0.168\pm0.003$ & --13.4 & --5.20 \\
~~~~Scaled Semi-major Axis & $a/R_{*}$ &
$56.22_{-1.66}^{+1.59}$ & $54.01_{-1.52}^{+1.78}$ & --3.9 & --1.33 \\
~~~~Orbital Inclination & $i$ (degrees) &
$89.77_{-0.14}^{+0.15}$ & $89.64_{-0.08}^{+0.07}$ & --1.4 & --0.92 \\
~~~~Impact Parameter & $b$ &
$0.22_{-0.14}^{+0.15}$ & $0.35\pm0.07$ & +59.1 & +0.86 \\
~~~~Transit Duration & $T_{14}$ (hours) &
$4.67_{-0.10}^{+0.12}$ & $4.70\pm0.04$ & +0.6 & +0.25 \\
{~~~~Photometric Jitter, TESS} & $\sigma_{\rm{TESS}}$ (ppm) & 
$0.01_{-0.01}^{+5.62}$ & $0.02_{-0.02}^{+4.16}$ & +100.0 & $<$ 0.01 \\
{~~~~Dilution, TESS} & $D_{\rm{TESS}}$ & 
$1$ & $1.086\pm0.025$ & +8.6 & --- \\
{~~~~Photometric Jitter, Perkin} & $\sigma_{\rm{Perkin}}$ (ppm) & 
--- & $2360_{-260}^{+300}$ & --- & --- \\
{~~~~Photometric Jitter, Davey} & $\sigma_{\rm{Davey}}$ (ppm) & 
--- & $5090_{-340}^{+330}$ & --- & --- \\
{~~~~Photometric Jitter, ARCTIC} & $\sigma_{\rm{ARCTIC}}$ (ppm) & 
--- & $2010_{-70}^{+80}$ & --- & --- \\
{~~~~Dilution, ARCTIC} & $D_{\rm{ARCTIC}}$ & 
--- & $1.002\pm0.021$ & --- & --- \\
\sidehead{Planetary Parameters:}
~~~~Mass & $M_{p}$ ($M_{J}$) &
$0.66\pm0.07$ & $0.67\pm0.04$ & +1.5 & +0.14 \\
~~~~Radius & $R_{p}$  ($R_{J}$) &
$1.15_{-0.05}^{+0.04}$ & $0.99\pm0.03$ & --13.9 & --3.20 \\
~~~~Density & $\rho_{p}$ (g/\unit{cm^{3}}) & 
$0.54_{-0.10}^{+0.09}$ & $0.85\pm0.10$ & +57.4 & +3.44 \\
~~~~Surface Gravity & $\log{g_{p}}$ (cgs) & 
$3.095_{-0.056}^{+0.053}$ & $3.191_{-0.041}^{+0.039}$ & +3.1 & +1.81 \\
~~~~Semi-major Axis & $a$ (au) & 
$0.1587_{-0.0075}^{+0.0067}$ & $0.1525_{-0.0060}^{+0.0065}$ & --3.9 & --0.82 \\
~~~~Average Incident Flux & $\langle F \rangle$ ($10^8~\rm{erg/s/cm^{2}}$) & 
$0.039\pm0.003$ & $0.046_{+0.003}^{+0.004}$ & +17.9 & +2.33 \\
~~~~Equilibrium Temperature$^{a}$ & $T_{eq}$ (K) & 
$362\pm7$ & $378_{-7}^{+8}$ & +4.4 & +2.28 \\
\enddata
\tablenotetext{a}{The planet is assumed to be a black body.}
\tablenotetext{b}{Values from \citet{Canas2020_toi1899} used as the fiducial}
\end{deluxetable*}

\begin{figure*}[!hp]
    \centering
    \includegraphics[width=0.9\textwidth]{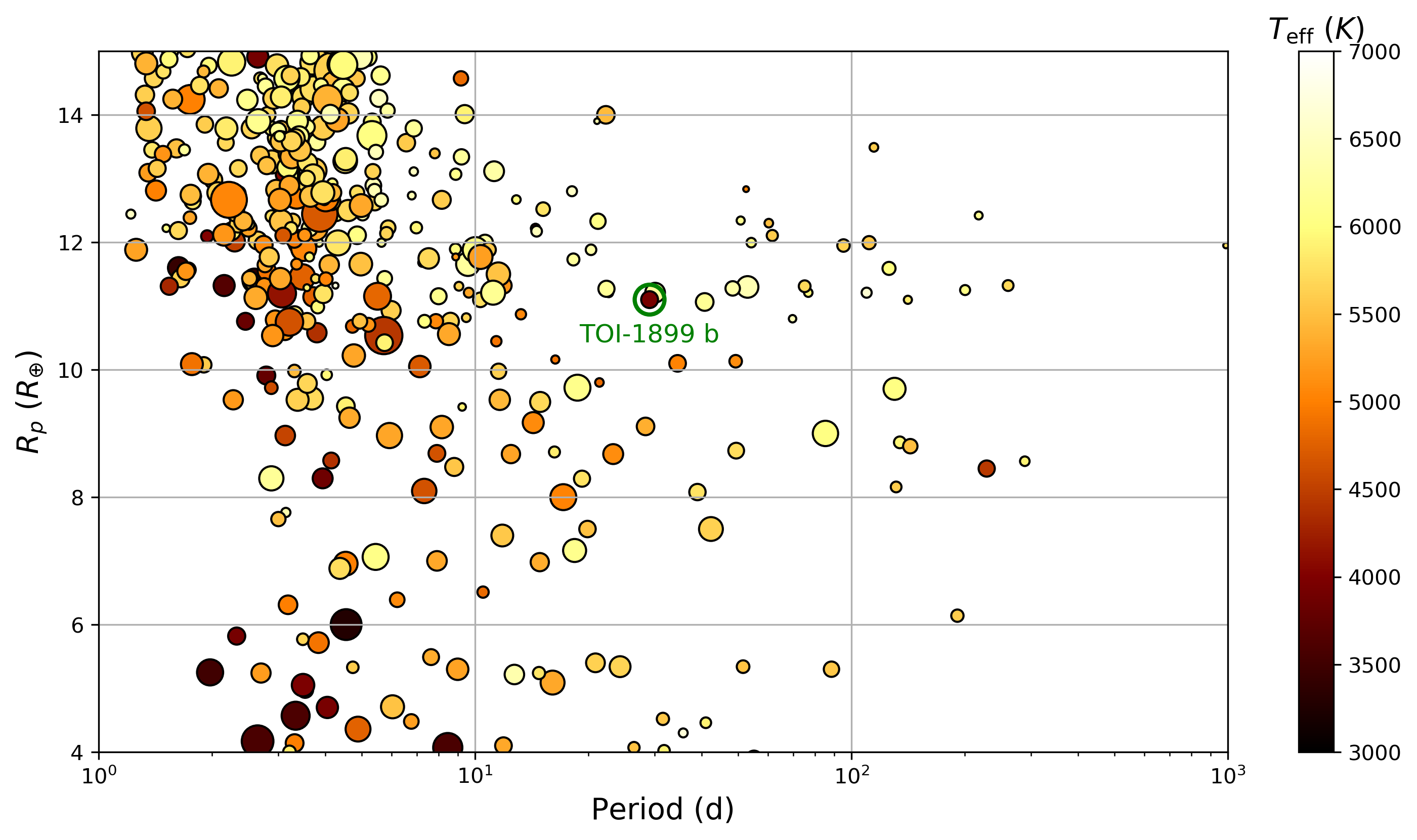}
    \includegraphics[width=0.9\textwidth]{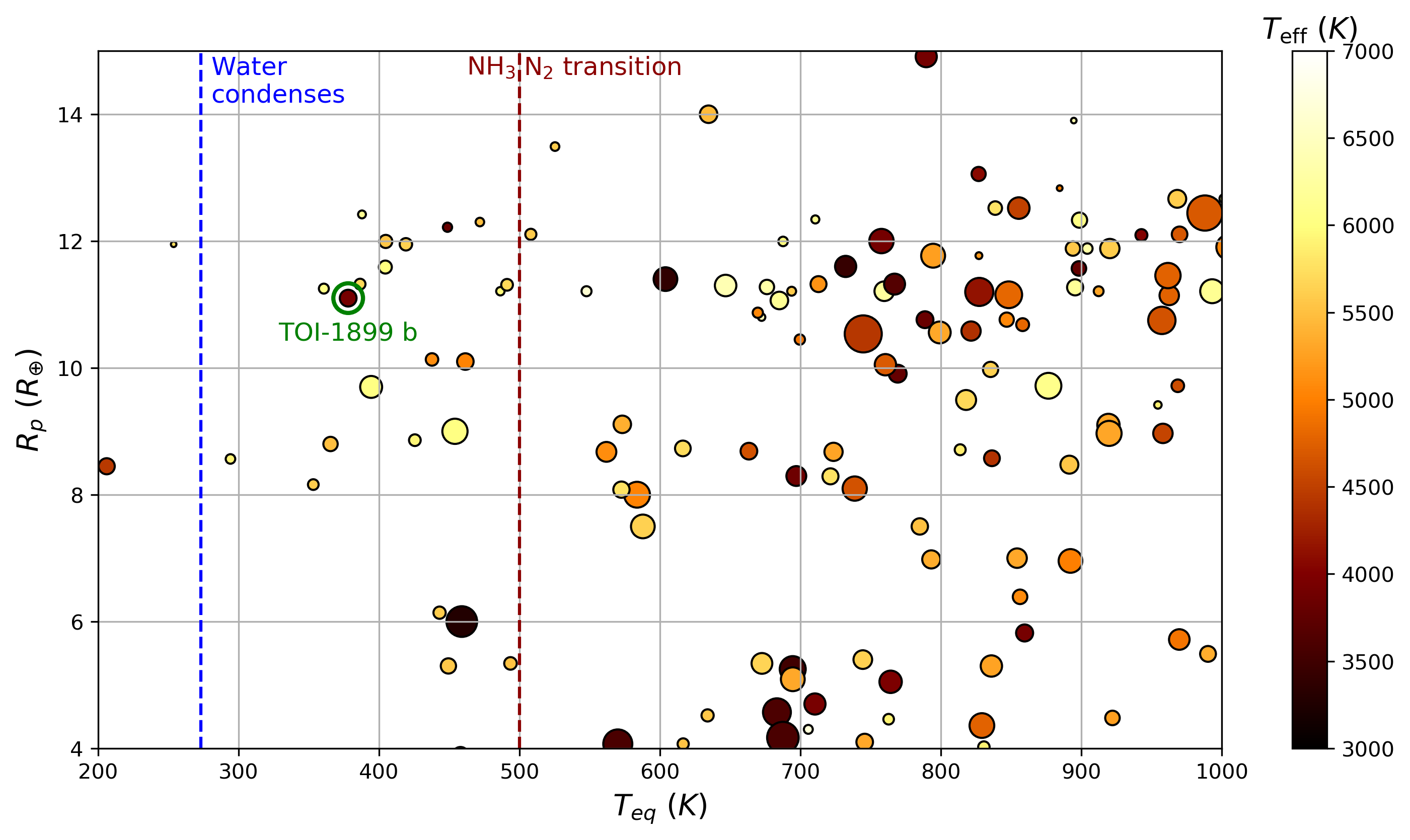}
    \caption{All transiting planets with masses measured to $>5\sigma$, plotted by planetary radius ($R_p$) vs. orbital period ($P$) (\textit{top}) and planetary radius ($R_p$) and equilibrium temperature ($T_{eq}$) (\textit{bottom}). TOI-1899~b is circled in green. Points are colored by the host star $T_{\rm{eff}}$ and scaled in size by the planet's Transmission Spectroscopy Metric \citep[TSM;][]{Kempton2018_tsm}. We also highlight the transition temperatures for water vapor and ammonia, species whose presence or absence in the atmospheres of cool planets like TOI-1899~b may hint at their formation and cooling history \citep{Fortney2020_atmochem}. TOI-1899~b is unique in $R_p$--$P$ space in being the only M-dwarf Jupiter with $P$ > 10~d, and in $R_p$--$T_{eq}$ space by having among the highest TSMs of Jovian-sized planets with $T_{eq}$ = 300--500~K, while also being the only one whose host star has $T_{\rm{eff}} \lesssim$ 4000~K. Data retrieved from the NASA Exoplanet Archive \citep{nasa_exoarchive_pscomp} on 2023 May 15.}
    \label{fig:Rp_vs_Teq}
\end{figure*}

% \pagebreak

\section{Discussion}
\label{sec:discussion}

We show in \autoref{fig:Rp_vs_Teq} that TOI-1899~b presents a unique opportunity to study a planet that spans multiple regions of currently unexplored parameter space. It belongs to the small but steadily growing population of gas giants orbiting M-dwarfs, though TOI-1899~b has by far the longest period of these planets, and is therefore the only one with an equilibrium temperature $<$~400~K. How this planet formed and why it seems to be the only known transiting Jovian planet around an M-dwarf with $P$ $>$ 10 days is unclear---characterizing its atmosphere could provide insights surrounding TOI-1899~b's formation and evolutionary history. %Moreover, TOI-1899~b provides a potential link between extrasolar hot/warm Jupiters and the cold Jupiter in our own solar system, as well as a link between Jupiters orbiting M-dwarfs to those around FGK stars. These multiple science cases make TOI-1899~b a compelling target for future transmission spectroscopy observations.

\subsection{Prospects for Atmospheric Characterization}
\label{sec:atmosphere}

\begin{figure*}[!ht]
    % \begin{minipage}[b]{0.58\linewidth}
    % \centering
    % \includegraphics[width=\textwidth]{spectrumfig2.pdf}
    % \end{minipage}
    % \begin{minipage}[b]{0.42\linewidth}
    % \centering
    % \includegraphics[width=\textwidth]{p_propagation.png}
    % \end{minipage}
    \centering
    \includegraphics[width=0.9\textwidth]{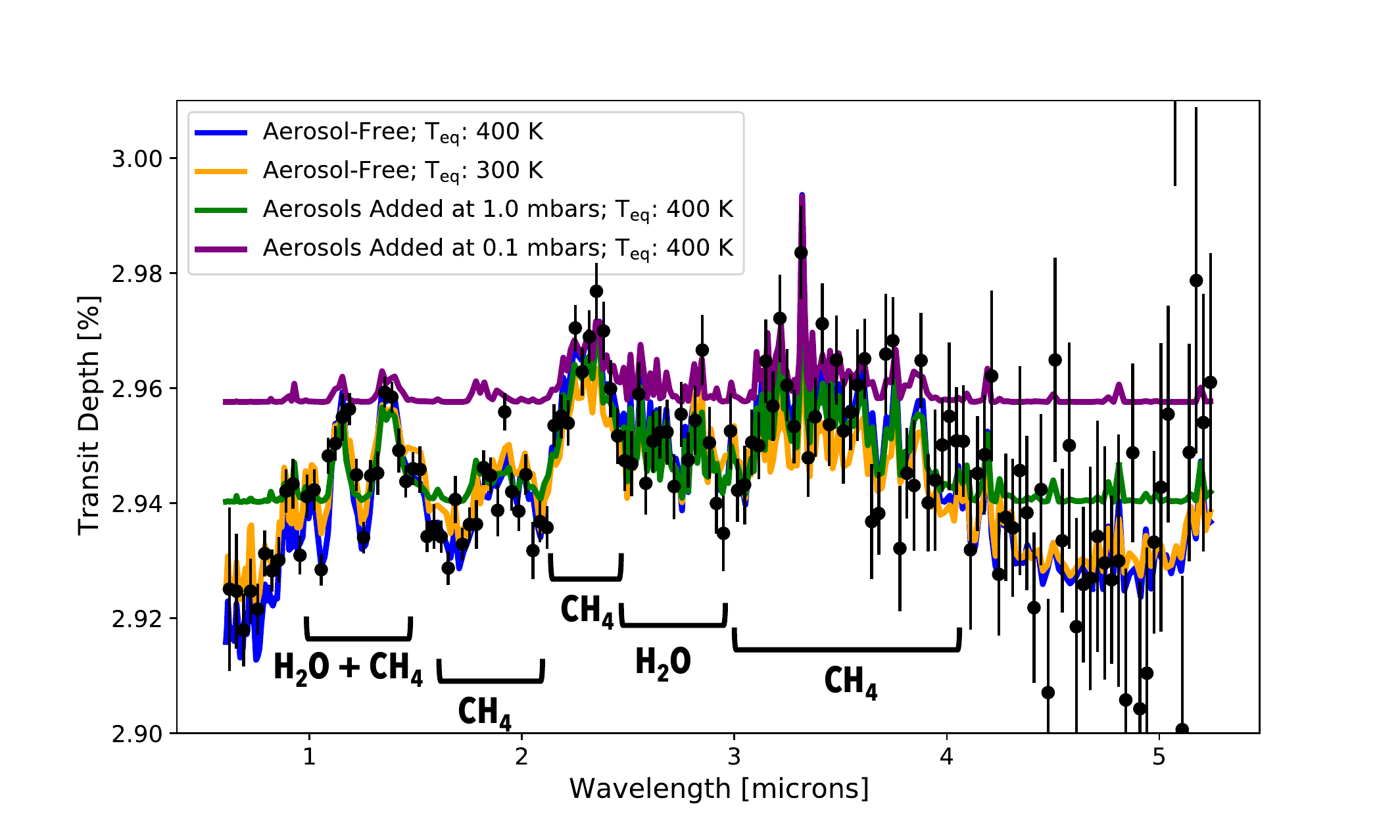}
    \caption{Simulated JWST/NIRSpec-Prism transmission spectrum (R $\sim$ 100) of TOI-1899~b created using a 10$\times$ Solar metallicity cloud-free atmosphere assuming a equilibrium temperature of 400~K and albedo of 0 (\textit{blue}) generated with \texttt{ExoTransmit} \citep{Exotransmit}. Errorbars are calculated using \texttt{PandExo} \citep{pandexo} assuming two observed transits of TOI-1899~b. The planetary spectrum is dominated by the absorption features of methane and water in this bandpass. We plot the same model atmosphere with the addition of an opaque cloud-deck (gray absorber) at pressure-levels of 1.0 mbars (\textit{green}) and 0.1 mbars (\textit{purple}) for comparison. As TOI-1899~b may have a similar atmosphere to Jupiter, we also consider a higher albedo of 0.5, leading to an equilibrium temperature of 320~K (300~K for \texttt{ExoTransmit} grid; \textit{orange}). The difference in temperature does not significantly change the spectra and should not impact JWST's ability to retrieve atmospheric abundances.}
    % \textit{Right}: Propagated uncertainties in the transit times of TOI-1899~b, for errors of 1$\sigma$ (darkest), 2$\sigma$, and 3$\sigma$ in the orbital period. Our updated ephemeris constrains the transit midpoint times to within $\pm$10~min (at 3$\sigma$) for the 5-year JWST primary mission, and is still viable for transit scheduling out to its maximum expected lifetime of $\sim$20 years ($\pm$40~min at 3$\sigma$).
    \label{fig:spectrum}
\end{figure*}

We calculate a Transmission Spectroscopy Metric \citep[TSM;][]{Kempton2018_tsm} of 47, putting TOI-1899~b into the third quartile for promising follow-up candidates according to \citet{Kempton2018_tsm}. However, as TSM is proportional to equilibrium temperature, it is no surprise that the cooler ($T_{eq} \approx$ 380~K, albedo $\approx$ 0) TOI-1899~b possesses a lower TSM than hotter planets of similar size. In comparison with other Jovian planets of similar $T_{eq}$ however, TOI-1899~b has among the \textit{highest} TSMs and a uniquely cool host star.

We simulate a cloud-free 10$\times$ Solar metallicity atmosphere model for TOI-1899~b using \texttt{ExoTransmit} \citep{Exotransmit} assuming an isothermal $P$-$T$ profile of 400~K, equilibrium chemistry (\autoref{fig:spectrum}, left), and uniform heat redistribution as TOI-1899b is unlikely to be tidally locked. Combining this model with \texttt{PandExo} \citep{pandexo}, we simulate a JWST/NIRSpec-Prism transmission spectrum based on observations of two transits. As shown in \autoref{fig:spectrum}, under these atmospheric conditions we could detect methane and water features with $>4\sigma$ confidence. Using the same model assumptions, we also simulate a 300~K planet with an albedo of 0.5, finding that the differences in the observed spectra are negligible and yield the same expected molecular features.

%We see in \autoref{fig:spectrum} that water and methane dominate TOI-1899~b's spectrum. Given its cool temperature, the presence of ammonia is also expected in TOI-1899~b's atmosphere \citep{moses.2013,fortney.2020}. Unfortunately, ammonia only possesses a weak absorption feature around 3.1-3.2 microns in the NIRSpec bandpass. MIRI observations at longer wavelengths (beyond 5 $\mu$m) are required to properly probe ammonia's absorption spectrum. However, we perform a quick retrieval of our simulated NIRSpec spectrum and find a tentative $\sim$2$\sigma$ detection for ammonia. Even though this is not a statistically strong measurement, combining the ammonia abundance with that of water and methane would provide a C/N/O ratio, a further constraint on TOI-1899~b's formation history not possible without JWST.

Aerosols (condensation clouds and photo-chemically created hazes) appear to be a common feature in most exoplanet spectra \citep{retrieval.review}. \citet{dymont.haze.trends} compared the presence of aerosols (derived from the water absorption feature strength in the Hubble/Wide Field Camera 3 bandpass) to a variety of stellar and planetary parameters for a set of 23 Neptunes and sub-Neptunes. Notably, the three planets observed by HST with temperatures between 300 and 500~K all show featureless spectra. While none of the planets analyzed by \citet{dymont.haze.trends} are Jovian, we nonetheless expect aerosols to be present in TOI-1899~b's atmosphere. We also include two cloudy models in Figure~\ref{fig:spectrum}---one with an opaque cloud deck (gray absorbers, wavelength independent) at pressure levels of 1.0 mbars and 0.1 mbars. For the 1.0 mbars cloudy model, we calculate a 9$\sigma$ significance of this model compared to a featureless flat line, with individual methane and water features detected to 3$\sigma$. These pressure levels were arbitrarily chosen to illustrate the precision JWST could achieve. Future work using photochemically-motivated haze models \citep[e.g. \texttt{VULCAN};][]{vulcan.citation} would provide new insights into haze formation of a cool Jovian world. %However, cloudy opaque aerosols at lower pressures/higher altitudes may completely mask all features even at the precisions obtained with JWST. 

While most relevant for rocky exoplanets with small-amplitude atmospheric features, the complicating effects of stellar contamination on abundance retrievals should not be ignored \citep[e.g.][who show that this effect could produce the water signature seen in the atmosphere of K2-18~b]{Barclay2021_k218_transpec}, especially for M-dwarf hosts with molecular features in their atmospheres \citep{LibbyRoberts2021_gj1132_transpec}. The presence of stellar surface inhomogeneities (e.g. starspots) can cause discrepancies in the transit depth as a function of wavelength \citep{Rackham2018_transpec}. Fortunately, TESS photometry of TOI-1899 shows no large-amplitude photometric modulations, and we show in Section \ref{sec:activity} that activity indices in both HPF and NEID spectra show no strong correlations with RVs, indicating that TOI-1899 is a relatively quiet and low-activity star. Therefore, any stellar contamination is likely to be minimal compared with the expected larger planetary signal.

Retrieving water and methane abundances with a JWST spectrum would provide a C/O ratio for TOI-1899~b---a potential measurement for constraining the disk location where TOI-1899~b formed based on snow-lines \citep[][]{oberg.snowlines, brewer.co.ratio}. Currently TOI-1899~b lies inside its star's snow-line. Understanding whether this planet formed \textit{in-situ} or migrated inwards would provide a key insight into the formation of the uncommon M-dwarf gas giants. We have constrained TOI-1899~b's mass and radius to $>$10$\sigma$ precision; thus, uncertainties in abundance retrievals will be dominated by JWST instrumental effects and the presence of aerosols, not from unconstrained planetary parameters \citep{batalha.precise.mass}. %In addition, our improved ephemeris enables confident scheduling of transit observations through the JWST primary mission and beyond.

%However, smaller hazy particles are not necessarily wavelength independent. \citet{kawashima.hu.haze} demonstrate that at wavelengths $>$2 $\mu$m the small haze particles become translucent as they are no longer large enough to absorb/scatter these wavelengths of light. Moreover, if these particles are tholins, they may also create their own absorption features, notably around 3 $\mu$m. In this scenario, it is possible that a high-altitude haze layer on TOI-1899~b may not mask all features in the NIRSpec bandpass.

\subsection{Tidally induced evolution}
\label{sec:tidal}

Temperate Jupiters such as TOI-1899~b would not be expected to experience significant tidal effects. However, since the host star is an M dwarf and the planet's orbital period is $\sim$29 days, the tidal evolution for this system can occur over timescales that are an order of magnitude larger than for short-period planets \citep{Dobs2004, Barker2009, Rodriguez2010, Alvarado2019} and four orders of magnitude larger than for ultra-short-period planets \citep{Brown2011, Wong2016, McCormac2019, Alvarado2021}.

\begin{figure}[htbp]
    \hspace{-16pt}\includegraphics[width=0.52\textwidth]{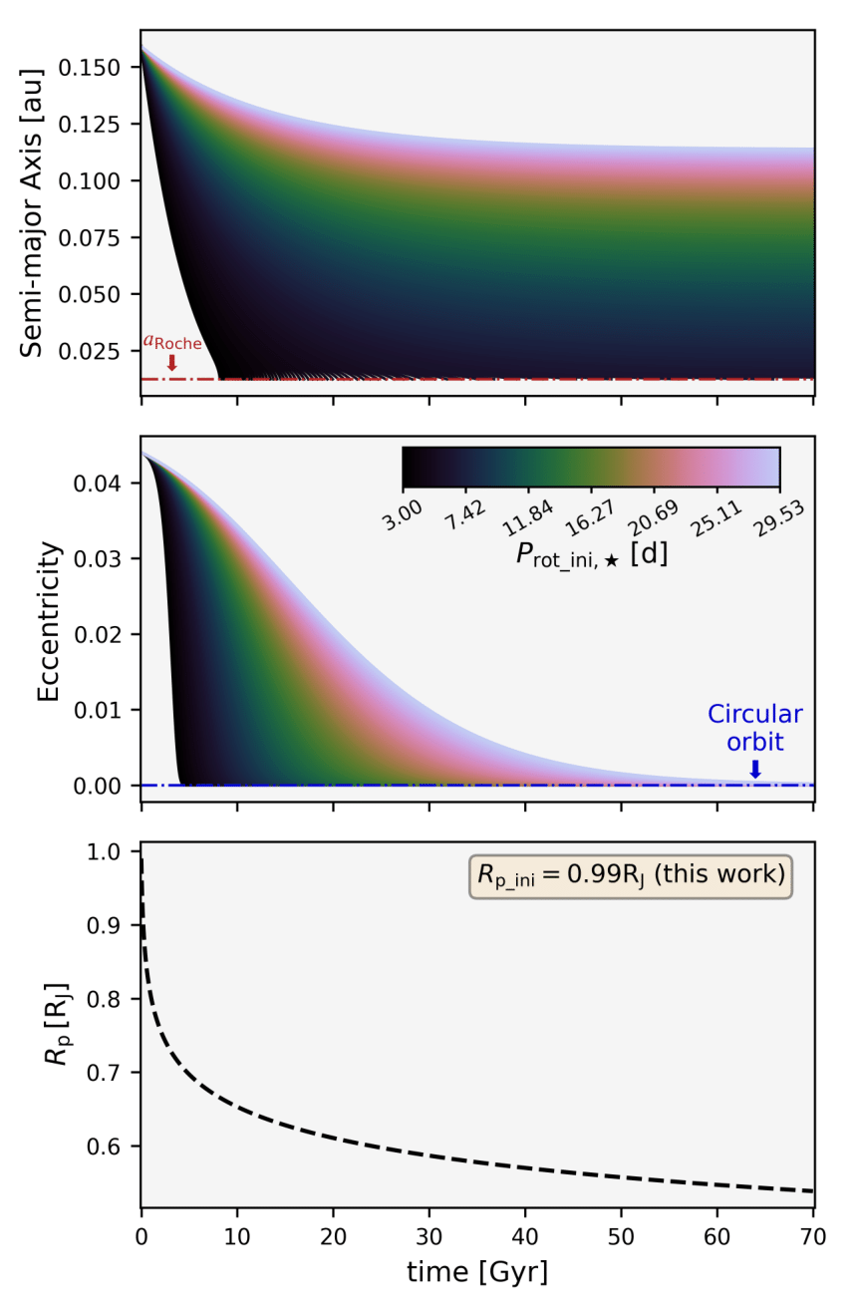}
    \caption{{\it Top}: Orbital semi-major axis, $a$, as a function of time. The colors represent different initial stellar rotation periods and the red dashed-dotted line stands for the system's Roche limit. {\it Middle}: Change in the eccentricity, $e$, for each $a$ on the top panel. {\it Bottom}: Evolution of planetary radius, $\Rp$, over the time-scales of tidally induced migration.}
    \label{fig:tidal}
\end{figure}

In this section we show that TOI-1899~b does not exchange a significant amount of angular momentum with its host star, and thus the stellar rotation will not spin up because of the presence of the planetary companion. In consequence, the tidal evolution of TOI-1899~b will take too long to be significant within $\sim13.5$ Gyr, unless the host star has a small initial rotation period, $\prot$ (\autoref{fig:tidal}, top panel).

Nevertheless, since tidal evolution depends on the gravitational forces between the interacting bodies, the study of this type of close-in giant planet orbiting an M dwarf, is necessary to analyze any deviation from the expected behavior of current theories of gravity, as has been done for other systems \citep{Matsumura2010}. Also, in some cases tidal interactions can produce eccentric orbits, so studying these systems can help us construct the path for any future study of their atmospheric dynamics \citep{Leconte2015}.

\begin{figure}[htbp]
    \hspace{-8pt}\includegraphics[width=0.50\textwidth]{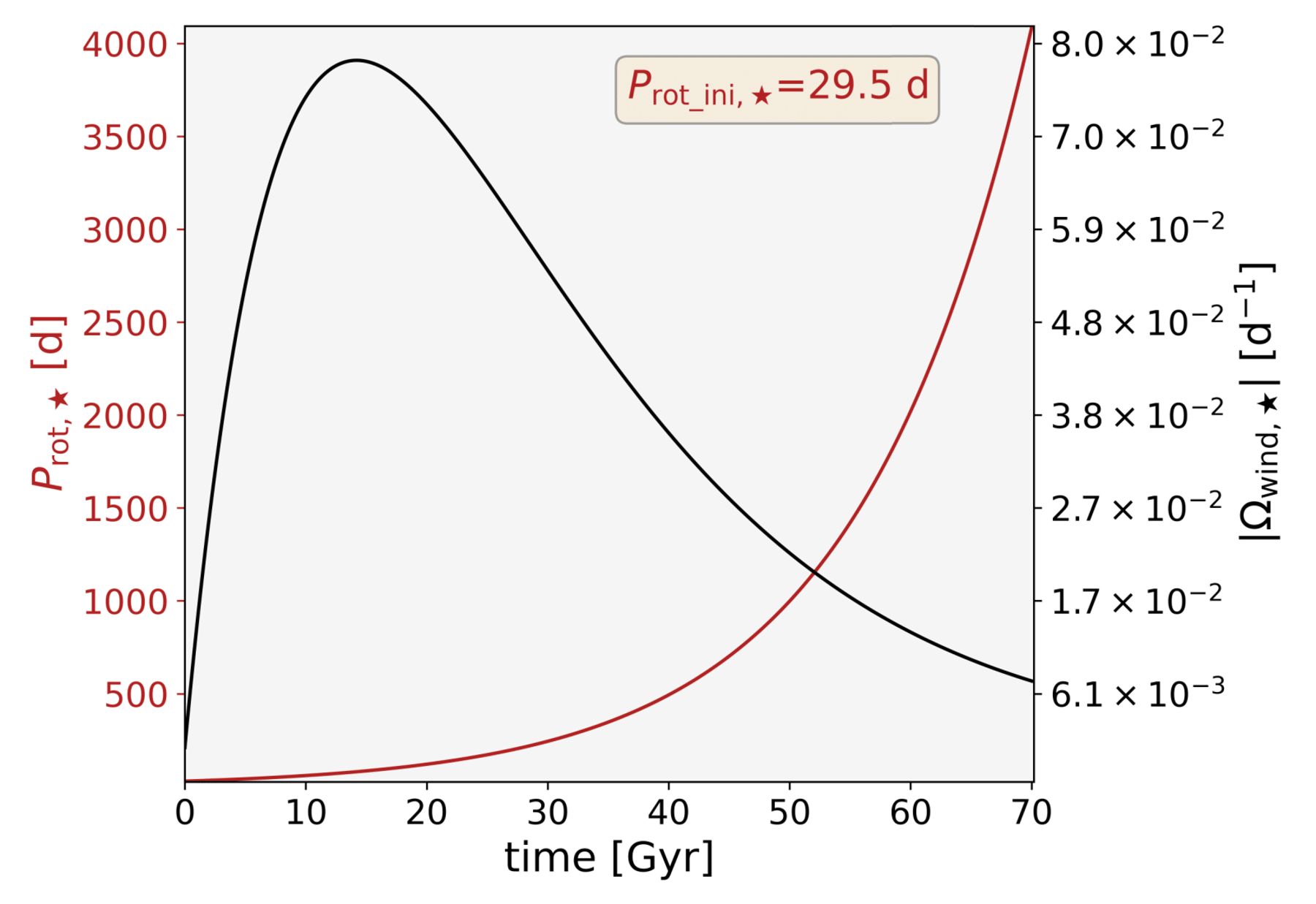}
    \caption{Stellar rotation period (\textit{red}, left-hand y-axis) and stellar wind braking rate (\textit{black}, right-hand y-axis), as a function of time. Both quantities were analyzed for a $\prot\sim29.5$ d and a star with half of its envelope contributing to the exchange of angular momentum (i.e. $\eps\gyrs=0.5$).}
    \label{fig:omegaperiod}
\end{figure}

Adopting the tidal formalism presented in \citet{Alvarado2021}, we study the orbital migration of TOI-1899~b until the stellar forces overcome the planet's self gravity at the Roche limit \citep{Roche1849}---where the planet starts being physically disrupted \citep{Guillochon2011}. As the planet decays in its orbit, we also integrate the stellar and planetary Tidal Dissipation Reservoirs\footnote{Represented by $k/Q$, the imaginary part of the second-order Love number which quantifies the system's tidal dissipation.} \citep[TDR;][]{Ogilvie2013, Guenel2014}, which provide insight into the exchange of angular momentum in the system. The top panel of \autoref{fig:tidal} shows the evolution of the planet semi-major axis for different initial stellar rotation periods, $\prot$. For $\prot\lesssim7$ d, TOI-1899~b undergoes orbital decay and crosses the Roche limit in less than 13.5 Gyr.

Assuming the observed orbital eccentricity as the initial value, we studied the forward tidal evolution of TOI-1899~b for a wide range of stellar rotation periods spanning up to $29.5$ d (see Section \ref{sec:rotation}). We found that orbital circularization timescales are only sensible for $\prot\lesssim7$ d (\autoref{fig:tidal}, middle panel), whereas longer $\prot$ led to the damping of the planet's eccentricity in timescales that are too long and prevent the planet from undergoing orbital decay.

For the planet, we assume Jupiter-like values for the core's rigidity (4.46$\times10^{10}$ Pa; \citealt{Lainey2009}) and for the interior mass and radius aspect ratios (i.e. $\alpha_\mathrm{p}=0.126$ and $\beta_\mathrm{p}=0.02$; \citealt{Mathis2015}). This yields a $\koQp\approx10^{-4}$ when the planet is tidally locked, which is unlikely for a 29-day orbital period. That, however, does not affect the overall evolution of the system because $\koQs$ are the overarching parameters that drive tidal dissipation. For the star, we adopt $\alpha_\star$ and $\beta_\star$ from \citet{Gallet2017} where the interior aspect ratios are presented for different types of stars. The adopted values give us $\koQs$ ranging from $10^{-7}$ to $10^{-9}$, for $\prot$ equal to 3 and 29.5 d, respectively.

In general, the star becomes less dissipative with time (i.e. small values of $\koQs$) due to the slow exchange of rotational (star) and orbital (planet) angular momentum, causing the stellar rotation period to increase (red line in \autoref{fig:omegaperiod}). As shown by the black line in \autoref{fig:omegaperiod}, during the first $\sim20$ Gyr of evolution the stellar wind braking rate increases, due mainly to the initial shrinking of the planetary orbit. However, the loss of material via stellar wind is the dominant mechanism for the evolution of the stellar rotation.

Given the initial distance of TOI-1899 b from the host star, we assume here that all tidal energy goes into circularizing the orbit and inducing orbital decay. Thus, during the tidal evolution of the system, we followed \citet{Fortney2007} to study the size contraction of TOI-1899~b. As shown by the bottom panel of \autoref{fig:tidal}, for migration times less than $20$ Gyr the planet contracts to $\sim35\%$ its initial radius, whereas for migration times larger than $20$ Gyr the planet's size decreases to $\sim42\%$ its original size. Although this affects $\koQp$, which contributes to the overall exchange of angular momentum (since it depends on the planet's physical structure), the overarching tidally-induced migration is commanded by the energy dissipated from the star's interior, which keeps constantly decreasing while the planet reaches a new orbital position. Such asymptotic decay with no crossing of the Roche limit occurs for $\prot\gtrsim7$ d.

It is worth mentioning that the tidal analysis performed here for TOI-1899~b has slightly different migration timescales depending on the amount of angular momentum driven by the envelope of the host star. This is represented by the product of $\gyrs$, the stellar gyration radius; and $\eps$, the fraction of the stellar envelope partaking in the exchange of angular momentum.\footnote{For a more complete description of these parameters and their values, see \citet{Dobs2004} and \citet{Alvarado2021}.} The product $\gyrs\eps$ will modify the times at which orbital decay occurs: in our analysis we set $\gyrs\eps=0.5$ (i.e. half of the stellar envelope exchanges angular momentum), but tidal migration could be slower for smaller $\gyrs\eps$, and vice versa. However, such variations are not significant for this system, as they are within the same order of magnitude as presented in \autoref{fig:tidal}.

Since $\porb\geq\frac{\prot}{2}$ in all the studied scenarios for $\prot$ in \autoref{fig:tidal}, we only calculate stellar dissipation through the excitation of inertial waves in the convective envelope. In addition, the mass of the planet is not sufficient for wave breaking to occur and thus internal gravity waves in radiative zones would not be fully damped by tidal forcing \citep{Barker2020}. The results presented here agree with those in \citet{Alvarado2022}, where the orbital evolution of TOI-1899 b induced by stellar tides is also minimum and not significant for most stellar rotation periods, with migration timescales being generally $\gtrsim10$ Gyr.

\section{Conclusion}

We have greatly refined the planetary parameters of the M-dwarf warm Jupiter TOI-1899~b, using 33 new RVs from the precision spectrographs HPF and NEID, and transit photometry from the ground and from three new sectors of TESS. Most notably, we derive a more precise period for this relatively long-period planet, $P = 29.090312_{-0.000035}^{+0.000036}$~d, and we find that TOI-1899~b is not as inflated as initially thought, with the original radius (derived from a single TESS transit) being $R_p = 1.15^{+0.04}_{-0.05}~R_J$, and our joint fit including ground-based data instead yielding a radius of $0.99 \pm 0.03~R_J$. We also find that the orbit is less eccentric than previously believed, with the new value of $e =$ $0.044^{+0.029}_{-0.027}$ consistent with zero at 2$\sigma$. The star TOI-1899 is activity-quiet over the past three years of RV observations and a slow rotator, with a tentative rotation period $\sim$29.5 days. Among the scarce handful of transiting gas giants orbiting M-dwarfs, TOI-1899~b is by far the coolest, with an equilibrium temperature of $T_{eq} \sim$ 380~K. With our new ephemeris allowing for extremely precise knowledge of transit times for many years, we encourage transmission spectroscopy studies of this unique planet with JWST, which may offer insights into the formation of gas giants around cool stars.

% why a good target?
% only cool Jupiter with known T_0, P, M, R to > 10 sigma
% large (enough) TSM -- fainter star but large transit depth -- mention scale height too
% transition between HJs and actual Jupiter

% T_eq ~ 350 K -- water condenses, ammonia present, C/O/N ratio gives disk environ (+ metallicity, mass vs. Z trend)
% host M dwarf -- compare impact on atmo vs FG dwarfs

% address haze

% address starspot contamination -- frequently an issue with M-dwarfs, but this star is very quiet

% FIGURE: simulated atmosphere observations with JWST (not full retrieval) [JESSICA]


\section*{Acknowledgments} % DONE

We thank the anonymous referee for valuable feedback which has improved the quality of this manuscript.

% funding acks
JAA-M is funded by the International Macquarie University Research Excellence Scholarship (`iMQRES’).
CIC acknowledges support by NASA Headquarters through an appointment to the NASA Postdoctoral Program at the Goddard Space Flight Center, administered by USRA through a contract with NASA and the NASA Earth and Space Science Fellowship Program through grant 80NSSC18K1114.
GS acknowledges support provided by NASA through the NASA Hubble Fellowship grant HST-HF2-51519.001-A awarded by the Space Telescope Science Institute, which is operated by the Association of Universities for Research in Astronomy, Inc., for NASA, under contract NAS5-26555.
Part of this research was carried out at the Jet Propulsion Laboratory, California Institute of Technology, under a contract with the National Aeronautics and Space Administration (80NM0018D0004).

% CEHW
The Center for Exoplanets and Habitable Worlds is supported by Penn State and the Eberly College of Science.
% ACI & Cyberlamp
Computations for this research were performed on the Pennsylvania State University’s Institute for Computational and Data Sciences Advanced CyberInfrastructure (ICDS-ACI), including the CyberLAMP cluster supported by NSF grant MRI-1626251.  This content is solely the responsibility of the authors and does not necessarily represent the views of the Institute for Computational and Data Sciences.
% PSU land grant
The Pennsylvania State University campuses are located on the original homelands of the Erie, Haudenosaunee (Seneca, Cayuga, Onondaga, Oneida, Mohawk, and Tuscarora), Lenape (Delaware Nation, Delaware Tribe, Stockbridge-Munsee), Shawnee (Absentee, Eastern, and Oklahoma), Susquehannock, and Wahzhazhe (Osage) Nations.  As a land grant institution, we acknowledge and honor the traditional caretakers of these lands and strive to understand and model their responsible stewardship. We also acknowledge the longer history of these lands and our place in that history.

% HPF
These results are based on observations obtained with the Habitable-zone Planet Finder Spectrograph on the HET. We acknowledge support from NSF grants AST-1006676, AST-1126413, AST-1310885, AST-1310875, AST-1910954, AST-1907622, AST-1909506, ATI 2009889, ATI-2009982, AST-2108512, AST-2108801 and the NASA Astrobiology Institute (NNA09DA76A) in the pursuit of precision radial velocities in the NIR. The HPF team also acknowledges support from the Heising-Simons Foundation via grant 2017-0494. 
% HET
The Hobby-Eberly Telescope is a joint project of the University of Texas at Austin, the Pennsylvania State University, Ludwig-Maximilians-Universität München, and Georg-August Universität Gottingen. The HET is named in honor of its principal benefactors, William P. Hobby and Robert E. Eberly. The HET collaboration acknowledges the support and resources from the Texas Advanced Computing Center. We thank the Resident Astronomers and Telescope Operators at the HET for the skillful execution of our observations with HPF. We would like to acknowledge that the HET is built on Indigenous land. Moreover, we would like to acknowledge and pay our respects to the Carrizo \& Comecrudo, Coahuiltecan, Caddo, Tonkawa, Comanche, Lipan Apache, Alabama-Coushatta, Kickapoo, Tigua Pueblo, and all the American Indian and Indigenous Peoples and communities who have been or have become a part of these lands and territories in Texas, here on Turtle Island.

% KPNO
Based on observations at Kitt Peak National Observatory, NSF’s NOIRLab, managed by the Association of Universities for Research in Astronomy (AURA) under a cooperative agreement with the National Science Foundation. The authors are honored to be permitted to conduct astronomical research on Iolkam Du’ag (Kitt Peak), a mountain with particular significance to the Tohono O’odham. 
% Contreras Fire (until Sept 2023)
Deepest gratitude to Zade Arnold, Joe Davis, Michelle Edwards, John Ehret, Tina Juan, Brian Pisarek, Aaron Rowe, Fred Wortman, the Eastern Area Incident Management Team, and all of the firefighters and air support crew who fought the recent Contreras fire. Against great odds, you saved Kitt Peak National Observatory.
% WIYN
Data presented herein were obtained at the WIYN Observatory from telescope time allocated to NN-EXPLORE through the scientific partnership of the National Aeronautics and Space Administration, the National Science Foundation, and the National Optical Astronomy Observatory. WIYN is a joint facility of the University of Wisconsin–Madison, Indiana University, NSF’s NOIRLab, the Pennsylvania State University, Purdue University, University of California, Irvine, and the University of Missouri.
% NEID
Data presented were obtained by the NEID spectrograph built by Penn State University and operated at the WIYN Observatory by NSF's NOIRLab, under the NN-EXPLORE partnership of the National Aeronautics and Space Administration and the National Science Foundation.  This work was performed for the Jet Propulsion Laboratory, California Institute of Technology, sponsored by the United States Government under the Prime Contract 80NM0018D0004 between Caltech and NASA. 
These results are based on observations obtained with NEID under proposals 2021B-0035 (PI: S. Kanodia) and 2022A-802765 (PI: C. Ca\~nas).
We thank the NEID Queue Observers and WIYN Observing Associates for their skillful execution of our NEID observations.

% APO / ARCTIC / diffuser
The ground-based photometry is based on observations obtained with the Apache Point Observatory 3.5-meter telescope, which is owned and operated by the Astrophysical Research Consortium. 
We acknowledge support from NSF grants AST 1907622, AST 1909506, AST 1909682, AST 1910954 and the Research Corporation in connection with precision diffuser-assisted photometry.

% ZTF
Based on observations obtained with the Samuel Oschin 48-inch Telescope at the Palomar Observatory as part of the Zwicky Transient Facility project. ZTF is supported by the National Science Foundation under Grant No. AST-1440341 and a collaboration including Caltech, IPAC, the Weizmann Institute for Science, the Oskar Klein Center at Stockholm University, the University of Maryland, the University of Washington, Deutsches Elektronen-Synchrotron and Humboldt University, Los Alamos National Laboratories, the TANGO Consortium of Taiwan, the University of Wisconsin at Milwaukee, and Lawrence Berkeley National Laboratories. Operations are conducted by COO, IPAC, and UW.

% SIMBAD /ADS
This research has made use of the SIMBAD database, operated at CDS, Strasbourg, France, and NASA's Astrophysics Data System Bibliographic Services. 
% EXOFOP
This research has made use of the Exoplanet Follow-up Observation Program (ExoFOP; \dataset[DOI:10.26134/ExoFOP5]{doi.org/10.26134/ExoFOP5}) website, which is operated by the California Institute of Technology, under contract with the National Aeronautics and Space Administration under the Exoplanet Exploration Program.
% NASA Exoplanet Archive
This research has made use of the NASA Exoplanet Archive, which is operated by Caltech, under contract with NASA under the Exoplanet Exploration Program.

% TESS
Some of the data presented in this paper were obtained from MAST at STScI. Support for MAST for non-HST data is provided by the NASA Office of Space Science via grant NNX09AF08G and by other grants and contracts. 
% TESS
This work includes data collected by the TESS mission, which are publicly available from MAST. 
The specific observations analyzed can be accessed via \dataset[DOI:10.17909/SBX7-VG73]{doi.org/10.17909/SBX7-VG73}.
Funding for the TESS mission is provided by the NASA Science Mission directorate.

% Gaia
This work presents results from the European Space Agency (ESA) space mission \textit{Gaia}. \textit{Gaia} data are being processed by the \textit{Gaia} Data Processing and Analysis Consortium (DPAC). Funding for the DPAC is provided by national institutions, in particular the institutions participating in the \textit{Gaia} MultiLateral Agreement (MLA). The \textit{Gaia} mission website is \url{https://www.cosmos.esa.int/gaia}. The \textit{Gaia} archive website is \url{https://archives.esac.esa.int/gaia}.


\facilities{HET (HPF), WIYN 3.5m (NEID), TESS, APO 3.5m (ARCTIC), \textit{Gaia}}


\software{
% \texttt{ArviZ} \citep{kumar_arviz_2019}, 
\texttt{AstroImageJ} \citep{Collins2017_aij}, 
% \texttt{astroquery} \citep{ginsburg_astroquery_2019}, 
\texttt{astropy} \citep{robitaille_astropy_2013, astropy_collaboration_astropy_2018},
\texttt{barycorrpy} \citep{Kanodia2018_barycorrpy}, 
\texttt{batman} \citep{Kreidberg2015_batman}, 
\texttt{celerite} \citep{foreman-mackey_fast_2017, foreman-mackey_scalable_2018},
% \texttt{corner} \citep{foreman-mackey_2016_corner},
\texttt{dynesty} \citep{Speagle2020_dynesty},
% \texttt{exoplanet} \citep{ ForemanMackey2021_exoplanet_zenodo, ForemanMackey2021_exoplanet_joss},
\texttt{EXOFASTv2} \citep{eastman_2019_exofastv2},
\texttt{ExoTransmit} \citep{Exotransmit},
\texttt{HxRGproc} \citep{Ninan2018_hpf_ccd},
\texttt{ipython} \citep{perez_ipython_2007},
\texttt{juliet} \citep{Espinoza2019_juliet},
% \texttt{lightkurve} \citep{lightkurve_collaboration_lightkurve_2018},
\texttt{matplotlib} \citep{hunter_matplotlib_2007},
% \texttt{MRExo} \citep{kanodia_mass-radius_2019},
\texttt{numpy} \citep{harris_numpy_2020},
\texttt{pandas} \citep{mckinney_pandas_2010},
\texttt{PandExo} \citep{pandexo},
% \texttt{PyMC3} \citep{Salvatier2016_probabilistic},
\texttt{radvel} \citep{Fulton2018_radvel},
\texttt{scipy} \citep{virtanen_scipy_2020},
\texttt{SERVAL} \citep{Zechmeister2018_serval},
% \texttt{starry} \citep{Luger2019_starry, Agol2020_analytic},
\texttt{TESS-SIP} \citep{hedges_2020_tess-sip},
\texttt{tglc} \citep{Han2023_TGLC},
% \texttt{Theano} \citep{the_theano_development_team_theano_2016}.
\texttt{thejoker} \citep{PriceWhelan2017_thejoker}.
}



% \clearpage
\bigskip

\appendix

\section{Lack of activity correlation with HPF RVs}
\label{app:hpf_activity}

\begin{figure*}[!p]
    \centering
    \includegraphics[width=0.8\textwidth]{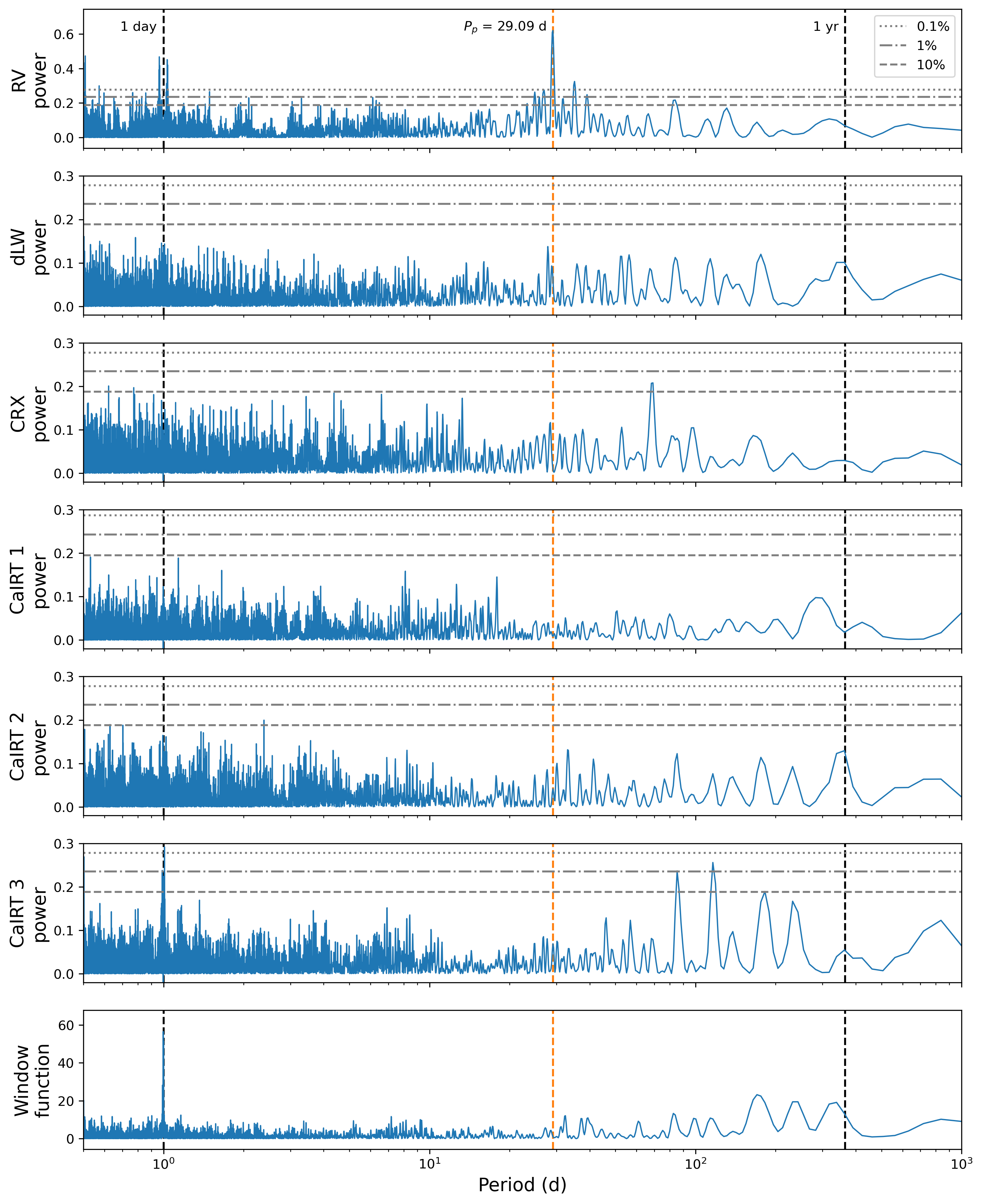}
    \caption{GLS periodograms of the HPF RVs, activity indicators, and window function, with the period of TOI-1899~b indicated by the vertical orange line. The FAP levels corresponding to 0.1\%, 1\%, and 10\% are also marked.}
    \label{fig:hpf_activity_pgram}
\end{figure*}

\begin{figure*}[!p]
    \centering
    \includegraphics[width=\textwidth]{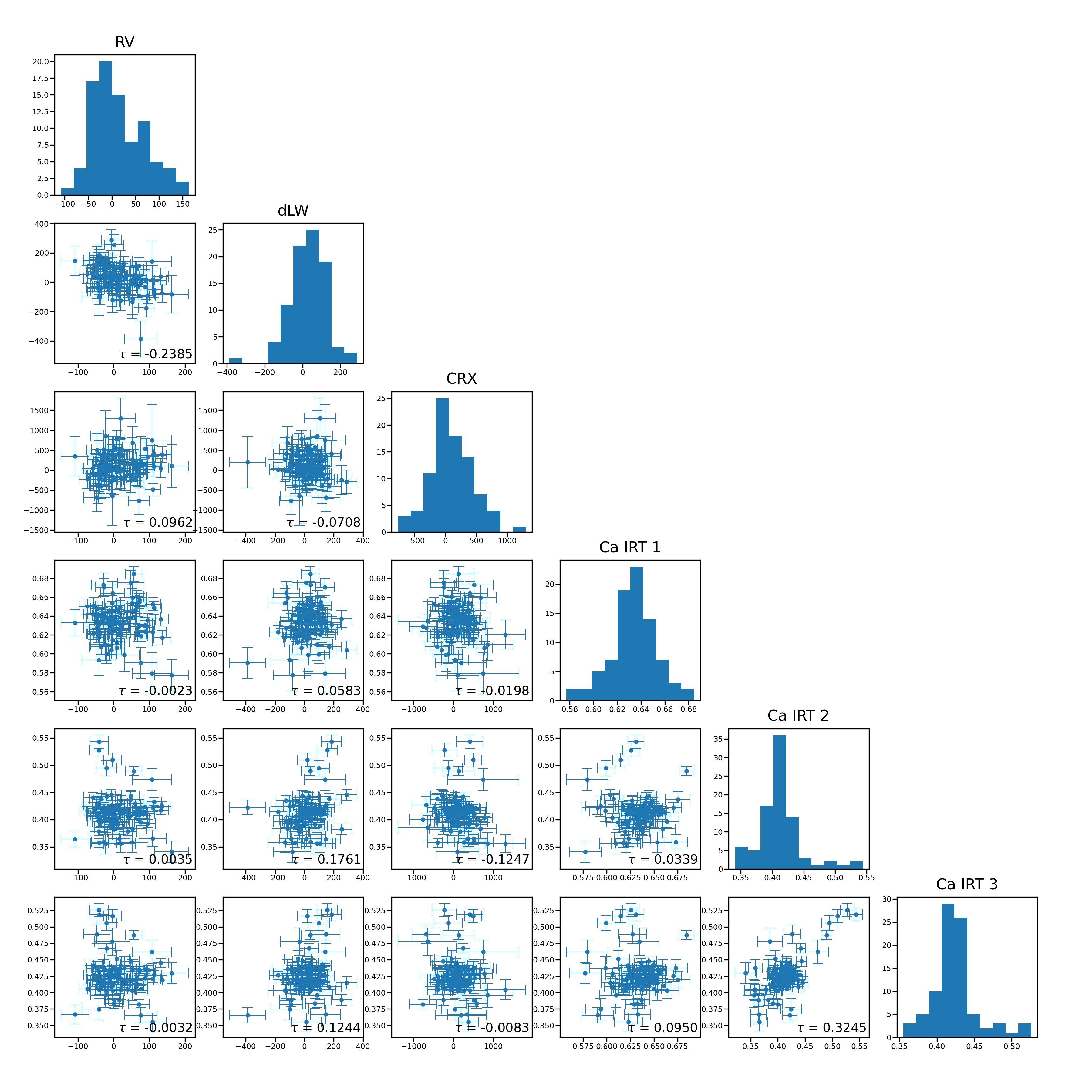}
    \caption{Corner plot of HPF RVs and activity indicators, with the Kendall $\tau$ coefficient included for each pair.}
    \label{fig:hpf_activity_corner}
\end{figure*}

As discussed in Section \ref{sec:activity}, we did not find any strong correlations between the HPF RVs  and the HPF activity indicators (dLW, CRX, and Ca IRT 1, 2, and 3 indices) using the Kendall rank correlation coefficient (Kendall $\tau$). We show the GLS periodograms in \autoref{fig:hpf_activity_pgram}, demonstrating that only the planetary RV signal greatly exceeds the false-alarm thresholds, and it is not associated with similarly-strong peaks in any of the activity indicators. \autoref{fig:hpf_activity_corner} further shows that, of the activity indicators, only dLW shows any correlation with RVs---but the correlation is weak ($\tau = -0.24$), and the corresponding periodogram peaks are far less significant than the RV signal. The data presented here are available as Data Behind Figure.

\section{Rotationally-modulated photometric variability}
\label{app:prot}

As discussed in Section \ref{sec:rotation}, we examined the publicly-available photometry from ASAS-SN in $V$ and $g$ (2015 Feb 24 to 2018 Nov 10, and 2018 Apr 12 to 2022 Dec 22, respectively), ZTF in $zr$ and $zg$ (DR17, which includes data up through 2023 Mar 9), and TESS (Sectors 14, 15, 41, 54, and 55) in an attempt to determine the rotation period of the host star TOI-1899. We present the GLS periodograms below.

\begin{figure*}[!p]
    \centering
    \includegraphics[width=0.9\textwidth]{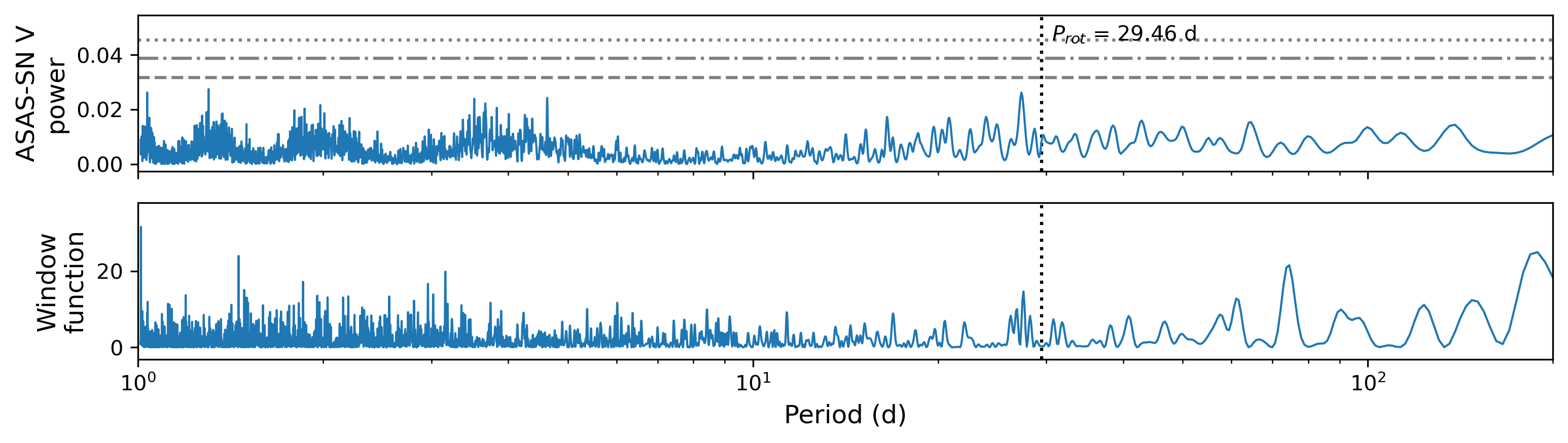}
    \includegraphics[width=0.9\textwidth]{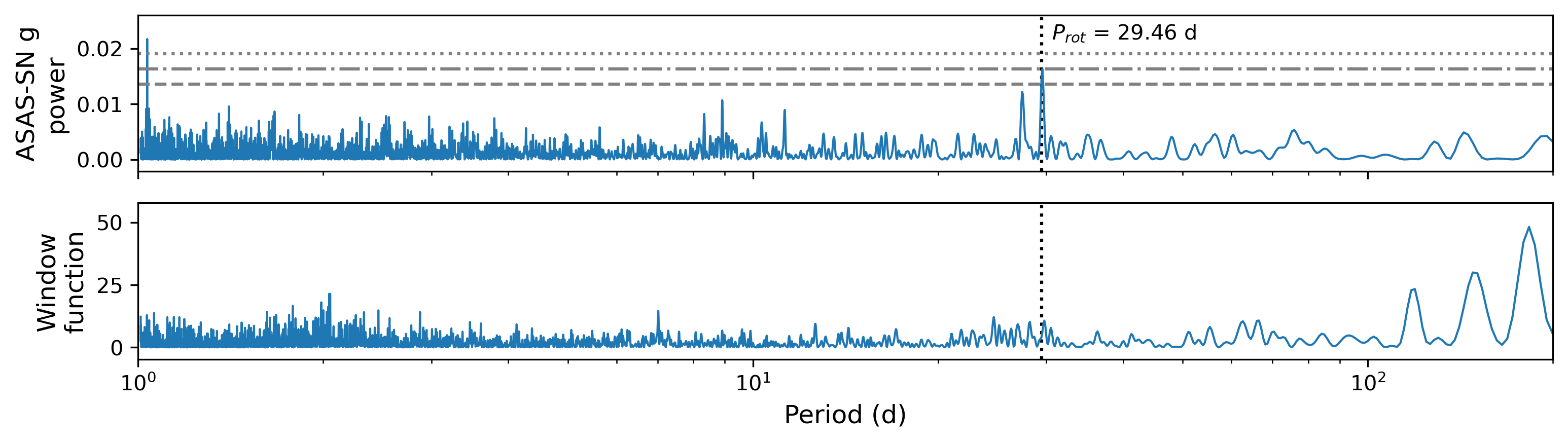}    
    \includegraphics[width=0.9\textwidth]{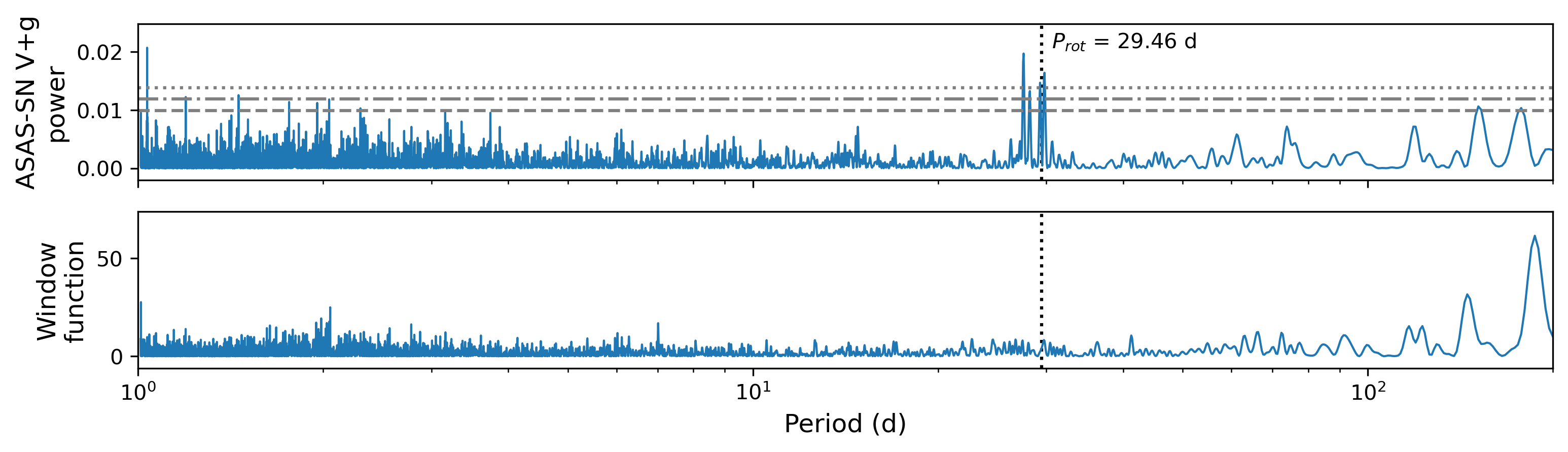}
    \caption{GLS periodograms of available photometry from ASAS-SN ($V$, $g$, and combined). We combine $V$ and $g$ in an attempt to more robustly detect long-period signals, since the two bands have little time overlap. FAP levels of 0.1\%, 1\%, and 10\% are marked as in \autoref{fig:hpf_activity_pgram}, and the tentative rotation period of 29.46~d is shown with a vertical dotted line.}
    \label{fig:pgram_asassn}
\end{figure*}

\begin{figure*}[!p]
    \centering
    \includegraphics[width=0.9\textwidth]{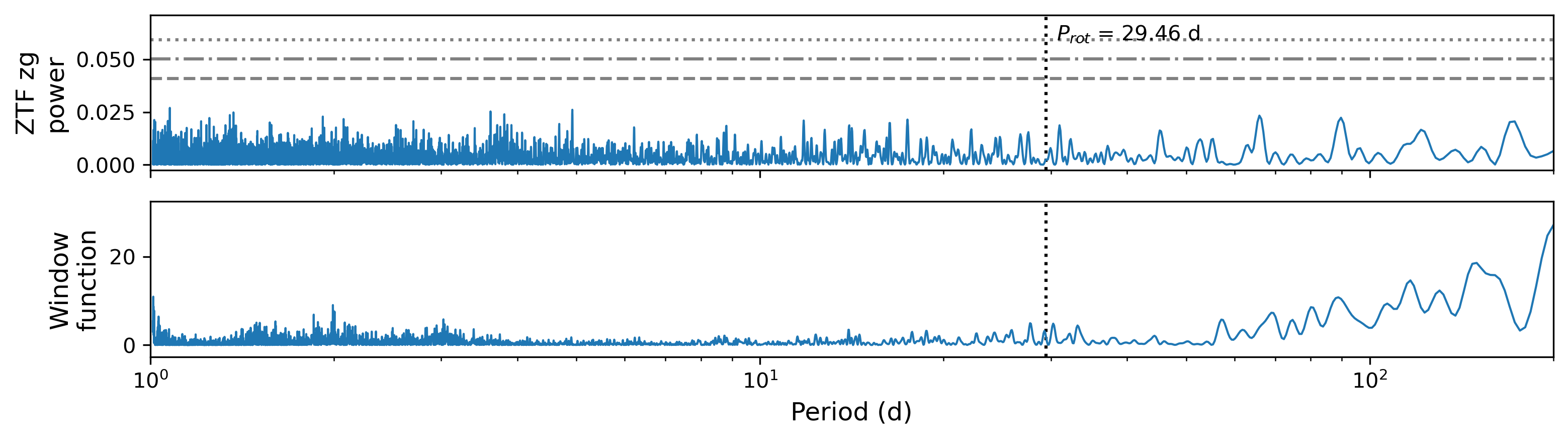}
    \includegraphics[width=0.9\textwidth]{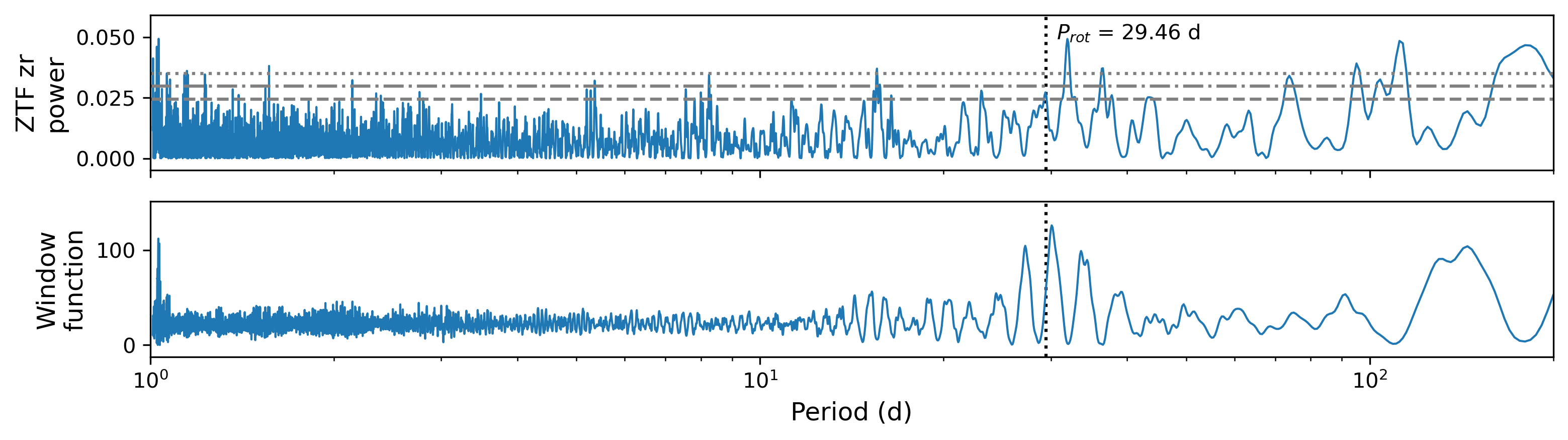}    
    \caption{GLS periodograms of available photometry from ZTF ($zg$ and $zr$).  Here we do not combine the bands as we did with the ASAS-SN data, since the $zg$ and $zr$ observations span the same timeframe.}
    \label{fig:pgram_ztf}
\end{figure*}

\autoref{fig:pgram_asassn} shows that combining the ASAS-SN $V$ and $g$ photometry to form a much longer baseline (see the top panel of \autoref{fig:prot_gp}) boosts the power of the peaks at $P \sim$ 27--29~d, which were detected at weak significance when the bands were considered separately. \autoref{fig:pgram_ztf} also shows peaks in ZTF $zr$ at $P \sim$ 31~d; however, with considerable power in the window function around these periods, aliasing effects are likely at play.

\begin{figure*}[!p]
    \centering
    \includegraphics[width=0.9\textwidth]{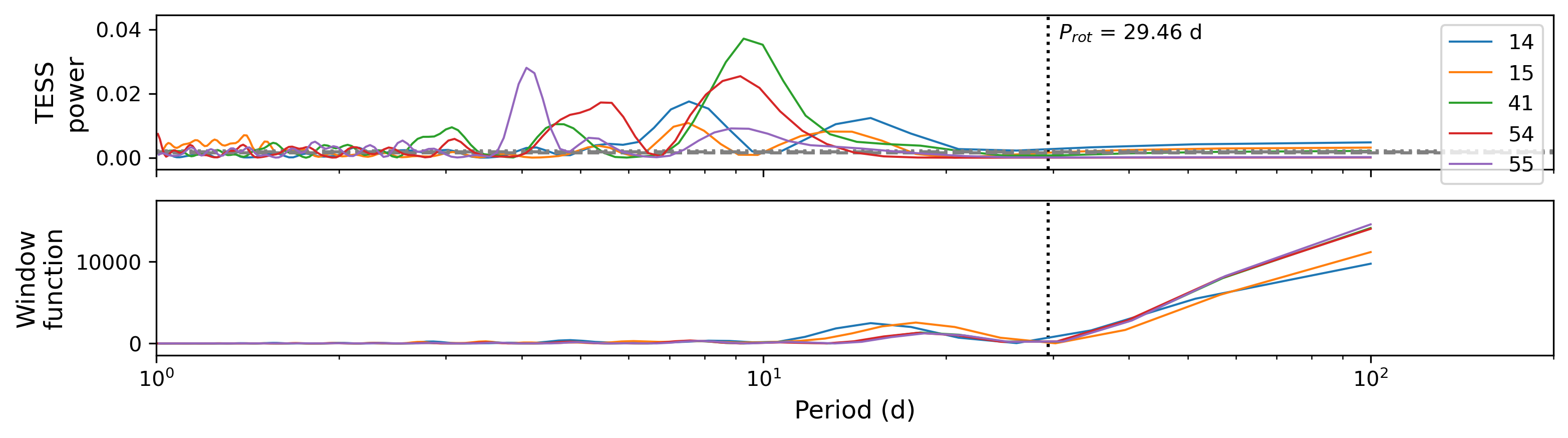}
    \includegraphics[width=0.9\textwidth]{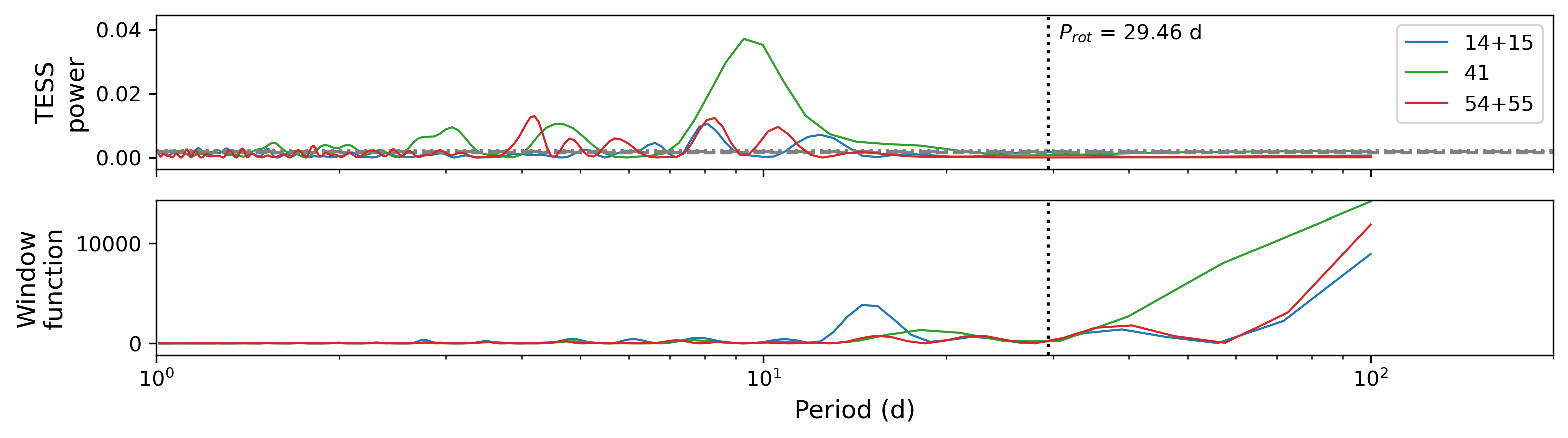}
    \caption{GLS periodograms of available PDCSAP photometry from TESS, with each sector separate (\textit{top}) and with consecutive sectors grouped together (\textit{bottom}). The $x$-axis is the same as previous plots for ease of comparison, but smoothing in the PDCSAP processing will suppress any signals $\gtrsim$ 10~d.}
    \label{fig:pgram_tess}
\end{figure*}

\begin{figure*}[htbp]
    \centering
    \includegraphics[width=0.9\textwidth]{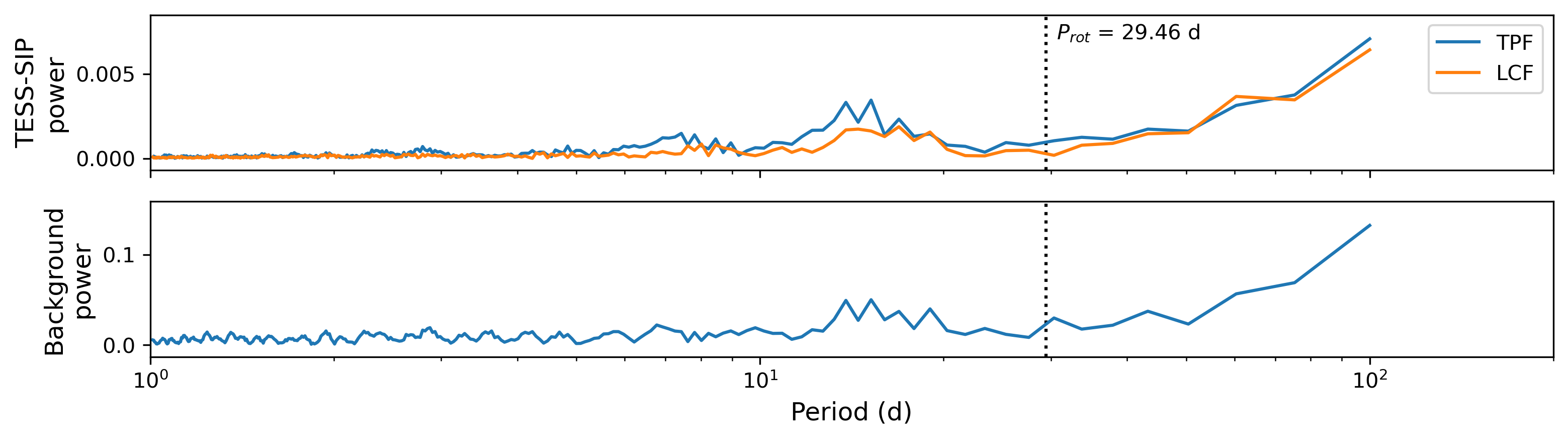}
    \caption{TESS-SIP periodograms computed from both the target pixel files (TPFs) and light curve files (LCFs), neither of which display the periodicity seen in the PDCSAP fluxes.}
    \label{fig:pgram_tess-sip}
\end{figure*}

In the TESS photometry (\autoref{fig:pgram_tess}), we find that the detected periods in the $P \sim$ 4--15~d range and the relative strengths of the peaks vary greatly from sector to sector---even consecutive sectors, e.g. Sectors 54 and 55. Furthermore, the systematics-insensitive periodogram from \texttt{TESS-SIP} (\autoref{fig:pgram_tess-sip}) reveals no strong peaks using either the target pixel files (TPFs) or the light curve files (LCFs), with the strongest signals correlated with background periodicity. Taken together, this suggests that the observed photometric variability in PDCSAP cannot be directly measuring the stellar rotation period; it may be an instrumental effect, or may trace some other form of low-level stellar activity.

\clearpage

% \clearpage

\bibliography{refs}{}
\bibliographystyle{aasjournal}



\end{document}